\documentstyle[prb,aps,preprint,tighten,eqsecnum]{revtex}
\begin{document}
\widetext

\draft

\date{29 August 1995}

\title{Random Matrix Theory of the Energy-Level
Statistics of Disordered Systems at the Anderson Transition}
\author{C.~M.~Canali}
\address{International Centre for Theoretical Physics, 34100
Trieste, Italy\\}

\maketitle

\begin{abstract}

We consider a family of random matrix ensembles (RME) invariant
under similarity transformations and described by the probability
density $P({\bf H})= \exp[-{\rm Tr}V({\bf H})]$.
Dyson's mean field theory (MFT) of the corresponding
plasma model of eigenvalues is generalized to the case of weak
confining potential, $V(\epsilon)\sim {A\over 2}\ln ^2(\epsilon)$.
The eigenvalue statistics derived from MFT are shown to deviate
substantially from the classical Wigner-Dyson statistics when $A<1$.
By performing systematic Monte Carlo simulations on the plasma model,
we compute all the relevant statistical properties of the RME with
weak confinement. For $A_c\approx 0.4$ the distribution function of the
energy-level spacings (LSDF) of this RME coincides in a large energy window
with the LSDF of the
three dimensional Anderson model at the metal-insulator
transition. For the same $A_c$, the
variance of the number of levels,
$\langle n^2\rangle -\langle n\rangle^2$,
in an interval containing $\langle n\rangle$ levels on average, grows
linearly with $\langle n\rangle$,
and its slope is equal to $0.32 \pm 0.02$, which is consistent
with the value found for the Anderson model at the critical point.
\end{abstract}\draft


\section{Introduction}

Random Matrix Theory (RMT) was introduced by Wigner\cite{wigner}
and Dyson\cite{dyson62}
to provide a statistical description of
the quantized energy levels of heavy nuclei, and since then
it has been applied
to a great variety of complex systems, quantum and classical\cite{mehta}.
Gorkov and Eliasheberg\cite{gorkov} suggested that
the Wigner-Dyson (WD) statistics, derived from RMT,
could be used to describe the energy levels
of small metallic particles at low temperature,
in connection with the study of their electromagnetic properties.
Here a statistical description
is made necessary by the presence of disorder and
irregularities in the shape of the particles.
For the case of disordered conductors one can resort to powerful
field-theoretical techniques, which have allowed Efetov\cite{efetov83}
and Altshuler and Shklovskii\cite{alt_shkl86} to show
analytically that
the WD
statistics are more than a simple phenomenological conjecture
and describe exactly
the local fluctuations
of the energy levels in metals in a certain regime.

The WD statistics are characterized by strong energy-level
correlations, giving rise to the phenomenon of
the level repulsion.
These are the correlations that typically exist among the
eigenvalues of a Gaussian Ensemble (GE) of random Hermitian
matrices ${\bf H}$, that is, an ensemble of matrices randomly distributed
with probability density $P({\bf H}) \propto \exp[-{\rm Tr} {\bf H}^2]$.

The GE's (and therefore the WD statistics) do not bear any
hint of
the spatial dimensionality $d$ of a physical system. Furthermore,
they are, by definition, invariant under similarity transformations
and thus there is no basis preference in them. This means that
they can be applied only to particular regimes of a physical system where:\,
1) all the normalized
linear combinations of the eigenstates have similar properties;
2) the dimensionality is, in some sense, irrelevant.

For a disordered electronic system this is just the ergodic
regime of the metallic state. In the metallic phase the eigenstates are
extended structurless objects. If we further assume that all
the relevant times are larger than
the ergodic time $\tau_D = L^2/D$,($D$ being the diffusin coefficient),
any diffusive particle
can completely and homogeneously
fill the total
sample volume $V= L^d$
during its trajectory (ergodic regime) and thus does not feel
the space dimensionality.
Alternatively, the ergodic time defines a natural energy scale,
$E_c = \hbar /\tau_D$ known as Thouless energy.
The ergodic regime, and therefore RMT, is valid within energy
intervals $\epsilon << E_c$, or within an interval containing a number
of levels
$N< E_c/\Delta \sim g$, where $\Delta$ is the mean level spacing.
The quantity $g$ is known as the dimensionless conductance in
units of $e^2/\hbar$. When $\epsilon >E_c$ the level statistics
depends on the dimensionality and is different from WD distribution
(Altshuler-Shklovskii regime).

The non-ergodic regime is never reachable in the metallic phase
in the thermodynamic limit, in any energy interval containing
a large but {\it finite} number of levels,
since the dimensionless conductance $g$,
diverges in the limit $L \to \infty$.
Therefore the WD statistics describe exactly
the energy-level correlations of the metallic state in
the $L\to \infty$ limit,
that exists for $d>2$ at relatively small disorder.

For $d=1,2$ no metallic state exists in the thermodynamic
limit, if any disorder is present, the system being always an insulator.
For $d=3$, with disorder increasing, the system
goes through the Anderson transition\cite{palee} to the insulating state,
where all states are localized.\cite{note0} The level statistics in these
situations obviously cannot be described by the a-dimensional, classical RMT
of the GE. In particular, such simple U(N)
invariant random matrix ensembles (RME) cannot
be an appropriate
description of an Hamiltonian matrix whose eigenvectors undergo the
phenomenon of localization, since one can construct extended states
by a linear combination of localized states. In contrast, the proper
$P({\bf H})$ distribution should contain eigenvector-eigenvalues correlations
or a basis preference that exclude those unitary transformations which
would lead to the formation of such extended states.

Random banded matrices\cite{FM,kik} are an example of
non invariant RME and perhaps
more realistic for describing Hamiltonians of quasi-onedimensional
disordered electronic systems\cite{FM} and other quantum chaotic
systems.\cite{kik} Their statistical properties exhibit
a crossover from the WD to the Poisson statistics as a function
of the parameter $b/N$, ($b$ is the band width) which is similar to
what happens in quasi-onedimensional systems upon decreasing the
ratio $\xi/L$, $\xi$ being the localization length.

While the RMT description of the quasi-onedimensional disordered systems
can be provided by random banded matrices, an analogous description
of the energy statistics
near the critical point for $d>2$ is still missing. This problem has
recently become a very outstanding one, after
intense study, both
numerical\cite{shklovskii,hof93,hof94,evangelo,sear,zar1,zar2,montanbaux}
and analytical,\cite{KLAA,ArKL}
has shown that the spectral correlations at the
metal-insulator transition
are also universal
and very different form the WD and the Poisson
statistics.

There is not yet full consensus on the exact nature of the
critical energy-level statistics.
The main finding of the analytical treatment based on a scaling
analysis\cite{KLAA,ArKL} is the existence of a power-law decay of the
two-level correlation function with a non trivial critical exponent.
All numerical simulations show that at the critical point
the statistical fluctuations of the energy levels are scale invariant.
However while some authors\cite{hof94,evangelo,montanbaux} claim a
good agreement of their numerical results with
the analytical predictions, others\cite{shklovskii,sear,zar1,zar2}
suggest that the critical statistics
constitute more simply a ``hybrid'' between the WD and the Poisson
distributions without any nontrivial critical exponent.

The search of a RMT description of the critical statistics has
prompted the investigation of physically motivated RME which exhibit
non trivial deviations from the
WD statistics. One important generalization has been obtained
in Refs.~[\onlinecite{pich_shap}, \onlinecite{moshe}]
starting from the
GE and introducing a symmetry breaking term of the form
\begin{equation}
\label{ShE}
P({\bf H})\propto e^{- Tr {\bf H}^2}\,e^{-h^2 N^2 Tr([{\bf \Lambda},{\bf
H}][{\bf \Lambda},{\bf H}]^{\dagger})}.
\end{equation}
The $h$-dependent term breaks the $U(N)$ invariance and tends to align
${\bf H}$ with a symmetry breaking unitary matrix ${\bf \Lambda}$ thus
setting the basis preference. It was shown in Ref.~[\onlinecite{moshe}]
that even after averaging over ${\bf\Lambda}$ the implicit presence of
the symmetry breaking term causes a dramatic change
in the level correlations of the resulting ensemble, which exhibit
a cross-over from the WD to the Poisson statistics.

On the other hand, a lot of work\cite{Pichardreview}
has been devoted to the analysis
of $U(N)$ invariant generalizations of the GE, of the kind
\begin{equation}
P({\bf H}) \propto e^{-{\rm Tr}\;V({\bf H})}
\end{equation}

Such ensembles arise from a {\it Global Maximum Entropy Ansatz} of RMT
in which an information entropy is maximized by the
distribution.\cite{Pichardreview}
The function $V({\bf H})$ acts like a generalized Lagrange multiplier
and it is determined, e.g., by requiring
that the density of eigenvalues is some given function $\rho(\epsilon)$,
taken directly from the microscopic system being investigated:
\begin{equation}
\label{PHinv}
\langle {\rm Tr}[\delta(\epsilon - {\bf H})]\rangle = \rho(\epsilon)
\end{equation}

For a long time it was believed that the
the local statistical
properties of the eigenvalues of these ensembles are completely
independent of $V$ and identical to those of the GE.
This hypothesis,
so far supported only by numerical evidence, has been proved more
rigorously very recently \cite{brezin,weid} for a large class of functions
$V$. On the other hand, it has also been
demonstrated\cite{muttalib,blecken,RMpap1,RMpap2}
that there exists
another class of functions $V$, known as {\it weak confining potentials},
for which the eigenvalue statistics display very strong deviations from
the universal WD behavior. Such potentials are characterized by a very
slow asymptotic growth:
\begin{equation}
\label{dblV}
V(\epsilon) \sim A\ln^2|\epsilon|\;, \quad |\epsilon| \to \infty\;.
\end{equation}
It is important to emphasize that
this asymptotic behavior of $V$
has been inferred from numerical studies of {\it Random Transfer Matrix}
models for disordered conductors, through the Maximum
Entropy Ansatz.\cite{Pichardreview,transfermatrix1,transfermatrix2}
In that case one considers
the eigenvalues of some combination of the transfer matrix, which are
directly related to the conductance and become larger
and larger, namely less confined, when the disorder increases.

It has been shown\cite{muttalib,blecken,RMpap1,RMpap2} that the
local eigenvalue fluctuations of the RME
with confining potential (\ref{dblV}),
exibit a cross-over from the WD statistics to a more Poisson-like behavior
when the parameter $A$ is decreased.
Since it is a common belief that there is a connection between
the statistics of eigenvalues and eigenstates of a RME,
this breakdown of the WD universality seems
to contradict
the argument, (presented
above for the GE), that a $U(N)$ invariant ensemble cannot exhibit
Poissonian statistics, typical of localized states.
In Ref.~[\onlinecite{RMpap2}] it has been suggested that
Poissonian behavior in such RME is a remarkable
phenomenon due to the spontaneous breakdown of the $U(N)$ symmetry
at the transition from a power-law potential,
$V(\epsilon) \sim |\epsilon|^\alpha$,
to the logarithmic potential of Eq.~(\ref{dblV}) when $\alpha\to 0$.
A crucial point in reaching this conclusion was the observation
that the two-level correlation function of the invariant RME
with weak confinement and the one of the RME with symmetry breaking
are identical in a certain range of the parameters.

Having established that these generalized RME with soft confinement
belong to a new universality class, characterized by nonclassical
correlations
that interpolate from the WD to the Poisson statistics,
there remains the important question of
whether or not their properties are related to the critical energy level
statistics of the Anderson model.

In this paper we address this question through a careful study of the
RME with distribution given by
Eq.~[\ref{PHinv}]. Firstly we shall show that the local statistical
properties
of a generalized RME of this sort can be correctly and easily determined using
an extension of Dyson's functional derivative formalism for the
corresponding one dimensional plasma model for the eigenvalues.
This formalism has been used successfully by Beenakker\cite{beenakker93prl}
to investigate
the random-matrix theory of the transmission eigenvalues in disordered
conductors.
In particular we prove that a modified mean field theory (MFT) of
the plasma model is able to yield the nonuniversal asymptotic
behavior of the two-level correlation function for weak confinement.

Secondly, we perform extensive Monte Carlo simulations of the
onedimensional Coulomb plasma of the eigenvalues and calculate
all the relevant quantities that describe short-range and long-range
statistical properties of the RME. The comparison with the MFT and other
analytical results shows that this Monte Carlo method provides
very accurate answers for this problem and can be used to study
more complicated RME where no analytical results are known.

Finally we critically compare the level statistics of these RME with the
the results recently obtained for the $3D$ Anderson model at the
metal-insulator transition. Our analysis shows that, while the asymptotic
correlations of the RME do not agree with the analytical result of
Ref.~[\onlinecite{KLAA}],
two other statistical properties, namely the
distribution function of the level spacings and level number variance,
are remarkably close to those found numerically in exact diagonalizations
of the Anderson model.

The paper is organized in the following way. In Section~[\ref{sec2}]
we set up formalism and notation and review Dyson's derivation of the
effective plasma model for the eigenvalue distribution. In
Section~[\ref{sec3:level1}] we develop a MFT of this model,
generalized to the weak confining potentials. The MFT study of
the two-level correlation function is carried out in Section~[\ref{sec4}].
Section~[\ref{sec5}] is devoted to the study of Monte Carlo simulations.
In Section~[\ref{sec6}] we discuss and compare the results of RMT with the
the analytical and numerical results of the energy level statistics
of the Anderson model at the metal-insulator transition.
Summary and conclusions are presented
in Section~[\ref{sec7}].

\section{Eigenvalues Statistics and effective one dimensional plasma model}
\label{sec2}

We consider an ensemble of random $N\times N$ matrices $\bf H$.
The matrices $\bf H$ are supposed to represent, for example, the
Hamiltonian of a complex system, such as a quantum disordered conductor.
We take $\bf H$ to be either {\it real symmetric}, {\it Hermitian},
or {\it quaternion-real self-dual}.\cite{mehta}
This choice defines three possible ensembles
corresponding to three different physical systems: 1) systems with
time-reversal
and rotational invariance; 2) systems with broken time-reversal symmetry;
3) systems with time-reversal symmetry but broken rotational invariance.

According to the maximum entropy principle
mentioned in the previous section,\cite{Pichardreview} we will assume
that the probability
distribution for the RME is defined by the density
\begin{equation}
\label{PH}
P({\bf H}) = Z^{-1}e^{-{\rm Tr}\;V({\bf H})}
\end{equation}
where $Z$ is a normalization constant.
The volume element is
$d[{\bf H}] = \prod_{i\geq j} d H_{ij}$,
for real symmetric matrices, with obvious generalizations
for the other two cases.\cite{mehta}
The probability density $P({\bf H})\;d[{\bf H}]$ is evidently invariant under
{\it orthogonal},
{\it unitary} or {\it simplectic} transformations respectively,
according to the three possible choices of $\bf H$.
The three ensembles are therefore specified and denoted after
their internal symmetry:
Orthogonal Ensemble (OE), Unitary Ensemble (UE) and
Symplectic Ensemble (SE).\cite{dyson62}

The invariance of $P({\bf H})$ in Eq.~(\ref{PH}) implies that
different matrices
with the same eigenvalues have the same probability to occur
in the distribution.

One can take advantage of this property and obtain the joint probability
distribution (JPD), ${\cal P}(\{\epsilon_i\})$,
for the {\it eigenvalues} $\epsilon_i,\;
i= 1,2,\dots \; , N$ of the matrices ${\bf H}$. For this purpose
it is necessary to express
the various components
of ${\bf H}$ in terms of the $N$ eigenvalues $\epsilon_i$ and other
mutually independent variables $p_j$, which together with $\epsilon_i$
form a complete set.
The variables $p_j$ can be integrated out and
the final result for JPD is\cite{mehta}
\begin{equation}
\label{PE}
{\cal P}(\{\epsilon_i\}) = C_{N_{\beta}}\exp\left(-{1\over 2}\beta
\sum_i^N V(\epsilon)\right)\prod_{i<k}\; |\epsilon_i - \epsilon_k|^{\beta}
\end{equation}
where $\beta=1$ for the OE, $\beta=2$ for the UE and $\beta=4$ for the SE.
$C_{N_{\beta}}$ is such that ${\cal P}$ is normalized to unity.
The {\it universal} Jastrow factor
$\prod_{i<k}\; |\epsilon_i - \epsilon_k|^{\beta}$ comes form the
Jacobian of the variable transformation. It is universal
in the sense that 1) it is always present in the JPD of the eigenvalues,
whenever the
the initial RME distribution probability is of the form Eq.~(\ref{PH}),
2) it is independent of the particular choice of
$V({\bf H})$ and depends only on the symmetry of the ensemble.

The JPD of Eq.~(\ref{PE}) characterizes all the statistical properties
of the eigenvalues
of an invariant RME, with the internal symmetry
discussed above. It describes the so called energy-level statistics,
if ${\bf H}$ is the Hamiltonian of the system.

Following Dyson\cite{dyson72}, the JPD can be rewritten in the
following form:
\begin{mathletters}
\begin{eqnarray}
\label{boltz}
{\cal P}(\{\epsilon_i\}) = Z_{N_{\beta}}^{-1}
\exp[-\beta {\cal H}(\{\epsilon_i\})]\; \\
\label{h}
{\cal H}(\{ \epsilon_{n}\})= - \sum_{i<j} \ln|\epsilon_{i}-\epsilon_{j}|+
\sum_{i}V(\epsilon_{i}).
\end{eqnarray}
\end{mathletters}

The probability distribution (\ref{PE}) has the form of a Gibbs distribution
for a classical, one dimensional system of $N$ ``particles'' $\epsilon_i$,
described by the ``Hamiltonian'' ${\cal H}$. The symmetry parameter $\beta$
plays the role of the equilibrium ``temperature''.

These fictitious ``particles'', namely the eigenvalues $\epsilon_i$,
interact among each other through a pairwise logarithmic repulsion
at any energy scale.
The external one-body potential $V(\epsilon)$ keeps the
system confined.
$V(\epsilon)$ is the only quantity of the RME that can be related to
the microscopic parameters of the original physical model through
its {\it global} statistical property, namely the density of eigenvalues.
[see below]. The logarithmic repulsion does not depend on any
microscopic detail of the real system, its origin being completely
geometrical. From now on we will use freely this particle model
analogy and call ``particles'' the eigenvalues of the RME.

The {\it local} statistical fluctuations of $1\approx n << N$ eigenvalues
are conveniently described by the
$n$-level correlation functions, defined as
\begin{equation}
G_n(\epsilon_1,\dots , \epsilon_n) = {N!\over (N-n)!}
\int_{-\infty}^{\infty}\; \dots \int_{-\infty}^{\infty}
{\cal P}_N(\{\epsilon_i\})\; d\epsilon_1\dots d\epsilon_n\; ,
\end{equation}

By definition, $G_n(\epsilon_1,\dots , \epsilon_n)$ is the probability
of finding simultaneously any $n$ particles at positions
$\epsilon_1, \epsilon_2,\dots , \epsilon_n$, the positions of the
remaining $N-n$ remaining unspecified. $G_n$ are positive defined.
In particular $G_1(\epsilon)$
gives the density of particles at position $\epsilon$, and it will
be denoted by
\begin{equation}
\label{mean_density}
\rho(\epsilon) \equiv G_1(\epsilon)
\end{equation}

It is convenient to introduce the
$n$-level cluster functions or cumulants. The {\it normalized} $n$-level
cluster function is defined by in the usual way of statistical
mechanics. The first two cumulants are

\begin{equation}
Y_1(\epsilon) = {G_1(\epsilon)\over \rho(\epsilon)} =1\; ,
\end{equation}
\begin{equation}
Y_2(\epsilon_1, \epsilon_2) = 1 - {G_2(\epsilon_1, \epsilon_2)\over
\rho(\epsilon_1)\rho(\epsilon_2)}\; ,
\end{equation}

We shall always consider the case
of large $N >>1$. In this limit the normalized cluster functions
are very useful since they tend to definite limits when the variables
are written in the correct units. In
taking the $N\to \infty$ limit it is necessary to measure
the particle positions in terms of the mean level
spacing, $\Delta$. If
$\lim_{N \to \infty} \rho(\epsilon) =  \rho_0 = {\rm constant}$,
$\Delta = \rho_0^{-1}$ and the dimensionless variables are simply
\begin{equation}
s_i = \epsilon_i/\Delta
\end{equation}
On the other hand if $\lim_{N \to \infty} \rho(\epsilon) \neq {\rm constant}$,
we need to consider a rescaling of the $\epsilon_i$ with the local
density or the more complicated unfolding procedure
[see Sec.~\ref{unfoldingsp} ].

In any case,
the $s_i$ will form a statistical model for an infinite
number of particles with mean spacing equal to unity.
It is only when written in terms of these rescaled variables
that the local statistical
properties of the eigenvalues of different RME
can be meaningfully compared.

We now come to the discussion on the explicit form of the
potential $V(\epsilon)$ and how the statistics depend on it.
The case of the Gaussian Ensembles (GE), with potential
$v(\epsilon) = \epsilon^2$
is the only one for which exact solutions for the density,
two-point correlation function and other statistical properties have
been known exactly for a long time.\cite{mehta} They are usually referred
as Wigner-Dyson (WD) or classical statistics.
One can show that in the large $N$ limit
the particle density obeys Wigner's semicircle law, with a radius
proportional to $N^{1/2}$. The two-point cluster function, in a
region around the origin $\epsilon=0$, is translationally invariant,
$Y_2(s_1, s_2) = Y_2(r), \, r=|s_1 - s_2|$.
For $\beta=2$ (GUE) it has the famous form
\begin{equation}
\label{Y2GUE}
Y_2(r) = \left[{\sin(\pi r) \over \pi r}\right]^2
\end{equation}
Similar expressions hold for the other two GE's.\cite{mehta}

At small separations the correlation function
$G_2(r) = 1- Y_2(r)$ vanishes, due to the phenomenon
of level repulsion brought about by the logarithmic interaction
\begin{equation}
G_2(r) \sim r^{\beta}\;, \quad r<<1
\end{equation}

Although the {\it global} statistics (such as the density)
of the energy spectra of real
systems do not follow the semicircle law, the {\it local} statistics
of the level correlations of many chaotic and complex systems are
very well described by Eq.~(\ref{Y2GUE}). In particular Eq.~(\ref{Y2GUE})
describes the correlations in small metallic samples at low
temperature,\cite{efetov83} as well as the correlations
among a large but finite number of energy levels of a metallic
system in the thermodynamic limit.\cite{KLAA}

Untill recently it was believed that the form of the confining
potential could only affect the density of eigenstates but not their
local statistics, which would therefore be universal
and equal to the WD statistics of the Gaussian Ensembles. Such
universality has been indeed proved recently in a rigorous way,
for a large class of potentials which confine the system strongly,
that is when $V(\epsilon)$ increases faster than
$|\epsilon|$.\cite{brezin,weid}

In what follows we will consider two kinds of non Guassian potentials,
which confine the system weakly,
with the following asymptotic behavior:

{ }

\noindent 1) Power-law Potential\cite{RMpap1,chenmanning,eriksen}
\begin{equation}
\label{Vpl}
V(\epsilon) = {A\over 2}|\epsilon|^{\alpha}, \quad 0<\alpha <1,
\qquad |\epsilon| \to \infty
\end{equation}

\noindent 2) Squared Logarithmic Potential\cite{muttalib,RMpap2}
\begin{equation}
\label{Vdbl}
V(\epsilon) = {A\over 2}\ln^2|\epsilon|, \quad |\epsilon| \to \infty
\end{equation}

It is a legitimate mathematical interest to investigate these cases
and see if the WD universality is preserved. However these RME have
also a physical intererst, since they have been suggested by random
transfer matrix models of disordered conductors.
\section{Mean Field Theory (MFT)}
\label{sec3:level1}

We now consider a mean field theory (MFT) analysis
of the classical one dimensional plasma model
of eigenvalues ${\cal P}(\{\epsilon_i\})$, into which
the original probability density of RME has been mapped. We first
define a continuous limit of this model,\cite{dyson62,dyson72}
valid in the asymptotic limit
of large $N$, by introducing the particle
density $\rho(\epsilon) = \sum_i ^N \delta(\epsilon -\epsilon_i)$.
In this limit we will assume that the Coulomb gas is a classical fluid
with a continuous and smooth
macroscopic density.
By substituting this definition of $\rho(\epsilon)$ in Eq.~(\ref{h})
the Hamiltonian
${\cal H}\{ \epsilon_{n}\})$ becomes
an {\it Energy} functional, ${\cal H}[\rho(\epsilon)]$,
of $\rho(\epsilon)$
\begin{equation}
\label{Hrho}
{\cal H}[\rho(\epsilon)] = - {1\over 2} \int_{-\infty}^{+\infty} d\epsilon
\int_{-\infty}^{+\infty} d\epsilon' \rho(\epsilon) \rho(\epsilon')
\ln|\epsilon-\epsilon'| +
\int_{-\infty}^{+\infty} d\epsilon \rho(\epsilon) V(\epsilon)\; .
\end{equation}

When $N$ is large but finite, the assumption of a smooth density is
only an approximation. Consequently the first term of
R.H.S. of Eq.~(\ref{Hrho})
does not reproduce exactly the corresponding term of Eq.~(\ref{h}),
because it neglects the two-level correlations, that is, it allows the
presence of the ``charges'' $\rho(\epsilon)d\epsilon$ and
$\rho(\epsilon')d\epsilon'$ at separations $\epsilon- \epsilon'\to 0$.
Since the interaction $\ln|\epsilon- \epsilon'|$ is singular, this
approximation has an effect, albeit small because $N$ is large.
One can compute the
correction in a $1/N$ expansion to this term and see that it is of the
form\cite{dyson72}
\begin{equation}
\label{correctio}
\delta {\cal H}[\rho(\epsilon)] =
-{1\over 2} \int_{-\infty}^{+\infty} d\epsilon
\rho(\epsilon) \ln (\rho(\epsilon))
\end{equation}

It is convenient to introduce a grand canonical potential
$\Omega [\rho] = {\cal H}[\rho] -\mu{\cal N}[\rho]$,
where ${\cal N}$ is the particle number functional
${\cal N}[\rho] = \int \rho(\epsilon) d\epsilon$
and $\mu$ is the chemical
potential.
The average density of particle $\langle\rho(\epsilon)\rangle$
can be expressed
in terms of the
functional integral
\begin{equation}
\label{ave-density}
\langle \rho(\epsilon)\rangle = Z^{-1}\int \rho(\epsilon)
e^{-\beta \Omega[\rho]}{\cal D} \rho \; ; \ \   Z= \int
e^{-\beta \Omega[\rho]}{\cal D} \rho \; .
\end{equation}
The one-body potential $V(\epsilon)$ acts as a source term for
the field $\rho(\epsilon)$ and $\langle \rho(\epsilon)$ can be expressed as

\begin{equation}
\langle \rho(\epsilon) = (\beta Z)^{-1}\delta Z/\delta V(\epsilon)\; .
\end{equation}

Up to now we have only assumed the
existence of a smooth particle density $\langle \rho(\epsilon)$, necessary in
taking the continuous limit of the Coulomb plasma, which presumably
is a good approximation when $N$ is large.
The MFT, based on the continuous approximation, amounts to
neglecting any entropy fluctuations
about the average and
using the saddle-point approximation
in the integral equation for $\langle \rho(\epsilon)\rangle$.
The MFT equation obeyed
by $\langle \rho(\epsilon)\rangle_{\rm MF}$ is therefore\cite{dyson72}
\begin{equation}
\label{MF_density}
\int d\epsilon'
\langle\rho(\epsilon')\rangle_{\rm MF}\,\ln|\epsilon-\epsilon'|=
V(\epsilon)-\mu ,
\end{equation}
where the ``Lagrange multiplier''
$\mu$ is to be found from the normalization
condition $\int \langle\rho(\epsilon)\rangle_{\rm MF}\, d\epsilon=N$.
The MFT interpretation of this equation is very natural, since
it represents the condition of mechanical equilibrium for the
charge density $\langle\rho(\epsilon)\rangle_{\rm MF}$ subject to the
external potential $V(\epsilon)$.
Such a MFT approximation
completely disregards the entropy part
${\cal S}[\rho] = -\int d\epsilon
\langle\rho(\epsilon)\rangle\ln \langle\rho(\epsilon)\rangle$
in the {\it Free-Energy} functional,
${\cal F}[\rho] = {\cal H}[\rho] -T{\cal S}[\rho]$,
and is exactly applicable only for
$\beta=\infty$. However, the long-range
nature of the pair-wise interaction in Eq.~(\ref{Hrho}) makes the MFT
approximation valid in the bulk of the spectrum even at finite
$\beta$, when $N \to \infty$.
Indeed Dyson\cite{dyson72} has calculated the first
correction to this equation
in a $1/N$ expansion and shown that the more accurate equation for
$\langle\rho(\epsilon)\rangle$ reads
\begin{equation}
\label{MF_density_N}
\int d\epsilon'
\langle\rho(\epsilon')\rangle\,\ln|\epsilon-\epsilon'|
+ {\beta -2 \over 2\beta}\ln\langle\rho(\epsilon)\rangle =
V(\epsilon)-\mu,
\end{equation}
The second term on the left-hand side of Eq.~(\ref{MF_density_N})
is the sum of two parts. The term proportional to $-T= -1/\beta$
comes clearly from the entropy contribution to the Free Energy.
The part {\it independent} of $\beta$ is generated in the passage to
the continuous limit as discussed above [see Eq.~(\ref{MF_density_N})].
The combination of these two terms acts as an entropy
contribution multiplied by an effective temperature, $T^{\star}(\beta)$
\begin{equation}
\label{Tstar}
T^\star (\beta)= \beta^{-1} - 1/2\; ,
\end{equation}
which vanishes for $\beta=2$.
We can write the Free Energy, at equilibrium,  of the system in the following
form
\begin{eqnarray}
\label{Fr}
F = &&-{1\over 2} \int_{-\infty}^{+\infty} d\epsilon
\int_{-\infty}^{+\infty} d\epsilon'
\langle\rho(\epsilon)\rangle\langle\rho(\epsilon')\rangle
\ln|\epsilon-\epsilon'| +
\int_{-\infty}^{+\infty} d\epsilon \langle\rho(\epsilon)\rangle
V(\epsilon)\nonumber\\
&&- T^{\star}(\beta)\int_{-\infty}^{+\infty}
d\epsilon[-\langle\rho(\epsilon)\rangle \ln\langle\rho(\epsilon)\rangle]
\; .
\end{eqnarray}
For the class of confining potentials of Eq.~(\ref{Vpl}) and
Eq.~(\ref{Vdbl}), one can
show\cite{dyson72} that
the relative
contribution of the correction to the MFT equation is of order
$N^{-1}\ln N$, and therefore vanishes in the thermodynamic limit.
For the time being we will neglect this correction and concentrate
on the MFT Eq.~(\ref{MF_density}).

\subsection{Solution of the eigenvalue density}
\label{sec3:level2a}

We now present the solution of the integral equation (\ref{MF_density})
for the particle density
$\rho(\epsilon)_{\rm MF}\equiv \langle\rho(\epsilon)\rangle_{\rm MF}$.
The solution of this equation, confined to the region $-D<\epsilon<D$, can
be found using the Cauchy
method \cite {Muskhelishvili} and is given by:
\begin{equation}
\label{MFS}
\rho_{\rm MF}(\epsilon)=\frac{1}{\pi^2}\sqrt{D^{2}-\epsilon^2}\, {\rm Re}
\int_{0}^{D}\frac{dV/d\xi}{\sqrt{D^{2}-\xi^2}}\,\frac{\xi
d\xi}{\xi^2-\epsilon_{+}^2},
\end{equation}
where $\epsilon_{+}=\epsilon+i0$ and the band-edge $D$ is to be found from
the normalization condition.
We will discuss separately the two cases of
a power-law and logarithmic potential.

\subsubsection{Power-law confinement,
$V(\epsilon) = {A\over 2}|\epsilon|^{\alpha}$}
\label{sec3:level3a}

The integral equation for $\langle\rho(\epsilon)\rangle_{\rm MF}$ becomes
\begin{equation}
\label{MFS_alpha}
\rho_{\rm MF}(\epsilon)={A\alpha\over \pi^2}\sqrt{D^{2}-\epsilon^2}\,
{\rm Re} \int_{0}^{D}\frac{\xi^{\alpha}}{\sqrt{D^{2}-\xi^2}}\,\frac{
d\xi}{\xi^2-\epsilon_{+}^2},
\end{equation}
We need to distinguish further between the cases $\alpha \geq 1$ and
$\alpha < 1$. For $\alpha \geq 1$, the integral in Eq.~(\ref{MFS_alpha})
is divergent in the thermodynamic limit $N\to \infty$, when the band edge
$D$ also goes to infinity. The main contribution to the integral is made by
the region $\xi \sim D$ and therefore for any fixed $\epsilon << D$, one
can neglect the $\epsilon$ dependence in the integrand of
Eq.~(\ref{MFS_alpha}). Then the mean level density tends to a {\it constant},
$\rho \to N^{1-1/\alpha}$. Thus we reach the important conclusion that,
for $\alpha \geq 1$,
there exists translational invariance in the $\epsilon$ space in the
$N\to \infty$ limit, exactly as in the case of the Gaussian ensemble.
Therefore for $\alpha \ge 1$ we have a condition
of strong confinement, and the corresponding local statistics
belong to the WD universality class in agreement
with Refs.~[\onlinecite{brezin},\onlinecite{weid}].
On the other hand, for $\alpha <1$, the integral in Eq.~(\ref{MFS_alpha})
is convergent even in the limit $D\to\infty$.  For finite $N$ ( and therefore
finite $D$), the integral can be calculated by transforming the original
contour of integration into a sum of two pieces,
$\Gamma= \Gamma_1 \cup \Gamma_2$, where $\Gamma_1$ is the negative imaginary
axis, and $\Gamma_2$ is a part of the positive real axis, $[D:\infty]$.
Since we are interested only
in the real part contribution, the integral over $\Gamma_2$
does not contribute
and we end up with
\begin{eqnarray}
\label{MFS_alpha_s}
\rho_{\rm MF}(\epsilon)&=&{A\alpha\over \pi^2}
\sin \left({\pi\alpha\over 2}\right)
\sqrt{D^{2}-\epsilon^2}\,
\int_{0}^{D}\frac{\xi^{\alpha}}{\sqrt{D^{2}+\xi^2}}\,\frac{
d\xi}{\xi^2+\epsilon^2}\\
\label{MFS_alpha_s2}
&=&
{A\alpha^2\over \pi D} C_{\alpha/2}\left({D\over 2}\right)^{\alpha}\,
\frac{\sqrt{1-z^2}}{|z|^{1-\alpha}}\;
F\left(\frac{1}{2},\frac{1+\alpha}{2};\frac{3}{2};1-z^2
\right)\; ,
\end{eqnarray}
\begin{equation}
\label{calpha}
C_{\alpha}=\frac{\Gamma(2\alpha)}
{\Gamma(\alpha)\Gamma(1+\alpha)}= {1\over 2} + O(\alpha^2)\; ,
\end{equation}
where $z=\epsilon/D$ and
$F(a,b;c;x)$ is a hypergeometric function. The band edge is
found from the normalization condition to be:
\begin{equation}
\label{D}
D=2\left(\frac{N\Gamma^{2}(\alpha/2)}{2A\Gamma(\alpha)}
\right)^{1/\alpha}
\end{equation}
The limiting function
$\rho_{\rm MF}^{\infty}(\epsilon) = \lim_{N\to \infty}\rho_{\rm MF}(\epsilon)$
is immediately obtained from Eq.~ (\ref{MFS_alpha_s2})
by taking the limit $z = \epsilon/D \to 0$:
\begin{equation}
\label{rho_infty}
\rho_{\rm MF}^{\infty}(\epsilon)=
{A\alpha\over 2\pi}\tan\left({\alpha \pi\over 2}\right)
{1\over |\epsilon|^{1-\alpha}}
\end{equation}
Therefore, for $\alpha <1$ the mean density in the thermodynamic limit,
$\rho_{\rm MF}^{\infty}(\epsilon)$, does
not scale as a power of $N$ and, especially, is not translational invariant.
$\rho_{\rm MF}^{\infty}$ is divergent at $\epsilon =0$. However, as
we will show in Sec.~[\ref{sec5}], this is an artifact of the MFT solution
and the exact density is sharply peaked but finite at the center.
We can say that the system undergoes a ``phase transition'' at
$\alpha=1$ and the symmetry which is broken is the translational invariance
of the problem in the limit $N\to \infty$.

\subsubsection{Squared logarithmic confinement,
$V(\epsilon) = {A\over 2}\ln^2(|\epsilon|)$}
\label{sec3:level3b}

In order to solve the integral equation for the MFT density, it is
convenient to represent this potential as a limit of combinations
of the power-law potential:\cite{note00}
\begin{equation}
\label{dblog}
V(\epsilon)=\ln^{2}|\epsilon|=\lim_{\alpha\rightarrow
0}[\alpha^{-2}(|\epsilon|^{\alpha}-1)^{2}]
\end{equation}
The MFT density in this case is
\begin{eqnarray}
\label{log2_rho_s}
\rho_{\rm MF}(\epsilon)= {4A\over \pi D}\sqrt{1 - z^2}
\Bigg[ &&\left({D\over2}\right)^{2\alpha} C_{\alpha}
{F\left(\frac{1}{2},\frac{1}{2}+\alpha;\frac{3}{2};1-z^2
\right)
\over z^{1-2\alpha}}\nonumber\\
&&- {1\over 2}\left({D\over2}\right)^{\alpha}
C_{\alpha/2}
{F\left(\frac{1}{2},\frac{1+\alpha}{2};\frac{3}{2};1-z^2
\right)
\over z^{1-\alpha}}\Bigg]\; ,
\end{eqnarray}
where $D$ satisfies the equation
\begin{equation}
N = {2A\over \alpha}\left[ C_{\alpha}\left({D\over2}\right)^{2\alpha}
- C_{\alpha/2}\left({D\over2}\right)^{\alpha} \right]
\end{equation}
The expression for the MFT density for $N$ finite
takes a simpler expression after the limit $\alpha\to 0$ is taken
\begin{equation}
\label{log2p}
\rho_{\rm MF}(\epsilon)=
A{\arcsin{\sqrt{1 -\epsilon^2/D^2}}\over\pi|\epsilon|}
\end{equation}
The band edge is now an exponential function of $N$:
\begin{equation}
\label{log2pD}
D = 2\exp(N/A)
\end{equation}
The singularity at $\epsilon=0$ is again an artifact of the MFT solution.
On the other hand we will see that in the bulk of the spectrum the mean
density is accurately described by Eq.~(\ref{log2p}).

\section{The two-level Correlation function}
\label{sec4}

In this section we study the two-level correlation function
using the MFT developed in the previous section.
Some of these results have been already derived for $\beta=2$,
using the different method of the orthogonal
polynomials. \cite{muttalib,RMpap2}
MFT will allow to generalize
these results to
any value of $\beta$.

\subsection{MFT Integral equation for the two-level correlation function}
\label{sec3:level2c}

Within the continuos formalism of the previous section,
let us consider the following definition of the
density-density correlation function $R_2(\epsilon,\epsilon')$
\begin{equation}
\label{R2}
R_2(\epsilon,\epsilon') = {{\langle\rho(\epsilon)\rho(\epsilon')\rangle}
\over {\langle\rho(\epsilon)\rangle \langle\rho(\epsilon')\rangle}} -1
\end{equation}
Using the definition of $\rho(\epsilon)$ one sees that
$R_2(\epsilon,\epsilon')$
differs from the two-level {\it cluster} function
$Y_2(\epsilon,\epsilon')$
defined in Eq.~(\ref{Y2GUE}) of
Section I by the singular self-correlation
\begin{equation}
\label{Y2}
Y_2(\epsilon,\epsilon') =
{1\over \langle \rho(\epsilon)\rangle}\delta(\epsilon-\epsilon')-
R_2(\epsilon,\epsilon')
\end{equation}
The correlation function $R_2(\epsilon,\epsilon')$ can be easily
expressed in terms of a functional derivative of $\langle\rho(\epsilon)\rangle$
with respect to $V(\epsilon)$ in Eq.~(\ref{ave-density}). By
using the relation $\delta \Omega/\delta V(\epsilon) = \rho(\epsilon)$ one
obtains\cite{beenakker93prl}
\begin{equation}
\label{R2_der}
R_2(\epsilon,\epsilon') = - {\beta^{-1}\over \rho(\epsilon)\rho(\epsilon')}
{\delta \langle\rho(\epsilon)\rangle\over \delta V(\epsilon)}
\end{equation}

Within the MFT it is possible to write down an integral equation for
the two-particle correlation function.\cite{beenakker93prl}
By taking the functional derivative
$ \delta / \delta V(\epsilon) $ in Eq.~(\ref{MF_density}) and using
Eq. (\ref{R2_der}) one obtains
\begin{equation}
\label{ieqR2}
\int d\epsilon''
\rho(\epsilon'')_{\rm MF}\,\rho(\epsilon)_{\rm MF}\,
R_2(\epsilon,\epsilon'')\,\ln|\epsilon'-\epsilon''|= -\beta^{-1}
\delta(\epsilon-\epsilon')+\beta ^{-1}\delta \mu/ \delta V(\epsilon),
\end{equation}
This important equation is sometimes used to claim and justify,
within the approximation
of MFT, the
universality of the correlations in RMT, or rather their independence of the
potential $V(\epsilon)$. The argument is usually the following. In the large
$N$ limit one implicitly assumes that, at least in the region of interest,
the average density scales like $N$ and goes to a constant,
$\langle\rho(\epsilon)\rangle = \rho_0$.
In this case, since the two body-interaction is translational invariant,
the two-particle correlation function must be translational invariant
as well.
In particular, the variational derivative of the chemical potential, which
is the only term that depends explicitly on the confining potential and
is not translationally invariant, must vanish.\cite{KLplasma}
If we now introduce
dimensionless variables $s=\epsilon/\Delta$, rescaled by the mean level
spacing $\Delta= \rho^{-1}$, one obtains an equation completely
independent of $V(\epsilon)$.
\begin{equation}
\label{gauss_r2eq}
\beta\int ds R_2(s-s'')\,\ln|s'-s''| = -\delta(s-s')
\end{equation}
{}From this one concludes that the limiting correlation
function $R_2(\epsilon-\epsilon')$ cannot depend on
the choice of $V(\epsilon)$ and it is universal.
Its asymptotic behavior is easily found by solving
Eq.~(\ref{gauss_r2eq}) by Fourier transformation. The result is of course
the universal WD behavior
of the Gaussian Ensembles.
\begin{equation}
\label{gauss_r2}
R_2(s-s') \sim -{1\over \pi^2\beta}{1 \over (s-s')^2}
\end{equation}
This reasoning is crucially based on the assumption that the average density
tends to a constant in the thermodynamic limit so that it can be simply
rescaled away. But we have seen
that there are cases in which, if the potential is weak
enough, the density is not at all a constant in the $N\to \infty$ limit
and in fact it decreases steeply with $\epsilon$. The system is not
translationally invariant and we cannot carry out the same simple rescaling
as before.
Thus the proof of the universality of the correlations
is not applicable, and one may expect
$R_2(\epsilon,\epsilon')$ to be
different from the WD form.
It is often said that in this case MFT breaks down and its
equations become invalid. This statement is not completely correct.
The MFT still gives reasonable and in fact accurate
results for the particle density in the bulk of the spectrum
(as the comparison
with numerical simulations will show). Therefore
this theory should work also for the correlations, when
properly applied. Of course we can no longer disregard the presence
of a nonconstant density, which is responsible of the deviations
from universality.
Below we suggest a simple
way out of this problem through the notion of {\it spectrum unfolding}.
\subsection{Spectrum Unfolding. Modified MFT equation for weak
confinement.}
\label{unfoldingsp}
On comparing the correlation functions of different ensembles it is
necessary to choose energy units such that the mean level spacing is
equal to one. This is trivial when the average density is a constant
$\rho$, since in that case $\Delta= \rho^{-1}$ and the convenient
variable is $s = \epsilon/\Delta =\epsilon \rho$.
The presence of a non uniform density in the thermodynamic limit
makes this linear rescaling impossible, even locally if the density is
a rapidly varying function of the energy, as in the case of the
logarithmic confinement.
This is a common situation in the study
of complex energy spectra, where one needs to subtract off the
unwanted effects of a non constant average density, in order to analyze the
fluctuations around the average density itself. This difficulty is handled
by the so called ``unfolding procedure'', namely the
introduction of a new variable
$s= s(\epsilon)$, in terms of which the density
becomes a constant. The variable that serves this purpose is the integrated
density of states
\begin{equation}
\label{unfolding}
s(\epsilon) = \int_0 ^{\epsilon}\langle\rho(\epsilon)\rangle\, d\epsilon\;.
\end{equation}
By particle conservation
\begin{equation}
\label{part_cons}
\langle \rho(\epsilon)\rangle\, d\epsilon =
{\langle\tilde{\rho}(s)\rangle}\, ds\;,
\end{equation}
where $ {\langle{\tilde \rho}(s)\rangle}$ is the average density in the
new variable $s$.
{}From Eqs.\ (\ref{unfolding}) and (\ref{part_cons}) it follows that
\begin{equation}
{\langle\tilde {\rho}(s)\rangle}=1\;, \ \ \ \ \ \langle s\rangle =1\; .
\end{equation}

The definition of the two particle correlation function in the unfolding
variables becomes
\begin{equation}
\label{R2_unf}
R_2(s,s') = {{\langle\rho(\epsilon_s)\rho(\epsilon_s')\rangle}
\over {\langle\rho(\epsilon_s)\rangle \langle\rho(\epsilon_s')\rangle}} -1\;,
\end{equation}
where $\epsilon_s$ is the inverse function of $s(\epsilon)$.
Using these definitions in Eq.~(\ref{ieqR2}) the integral equation
for $R_2(s,s')$ is\cite{noteonmu}
\begin{equation}
\label{ieqR2_unf}
\int ds''\, R_2(s,s'')\ln|\epsilon_{s''} -\epsilon_{s'}|=
-\beta^{-1}\delta(s-s')\; ,
\end{equation}

The MFT equation for $\langle{\tilde {\rho}(s)\rangle}$ reads
\begin{equation}
\label{ieqrho_unf}
\int_{\infty}^{\infty}ds' {\langle\tilde{\rho}(s')\rangle}
\ln|\epsilon_s -\epsilon_{s'}| = V(\epsilon_s) - {\mu}\; .
\end{equation}
Eqs.\ (\ref{ieqR2_unf}) and (\ref{ieqrho_unf}) together mean that
$\beta R_2(s,s')$ is the solving kernel of\cite{beenakker93prl}
\begin{equation}
\label{kern1}
\int_{\infty}^{\infty}ds\, \psi(s')\ln|\epsilon_s - \epsilon_{s'}|
= \varphi(s)  +
{\rm const}\; ,
\end{equation}
that is
\begin{equation}
\label{kern2}
\psi(s) = \int_{\infty}^{\infty}ds\, \beta R_2(s,s')\varphi(s')\; .
\end{equation}
The additive constant in Eq.~(\ref{kern1}) has to be chosen such that
$\int_{\infty}^{\infty}ds\psi(s) =0$ since the variations in
$\langle\tilde{\rho}(s)\rangle$ must occur at
constant $N$.\cite{beenakker93prl}

Eq. (\ref{ieqR2_unf}) is the integral equation for the correlation function
$R_2(s,s')$
in case of {\it weak confinement}.
{}From this equation we can see that, in contrast with Eq. (\ref{ieqR2})
valid for strong confinement, now
$R_2(s,s')$ depends implicitly on the original confining potential
through the unfolding variable $s(\epsilon)$. Thus it will not, in
general, be universal.

\subsection{Solution of the MFT integral equation}

The solving kernel of the integral equation (\ref{kern1}) must be
found separately
for each unfolding function $\epsilon(s)$, which in turn depends on the
confining potential.
We are interested in the correlations in the bulk of the spectrum,
where MFT works and thus we can use the MFT expression for the density in
order to
obtain the unfolding function.

We first consider the case of the power-law potential with
$0< \alpha <1$, which is the simplest case of weak confinement where,
in principle, non classical correlations can arise.
Using Eqs.~(\ref{rho_infty}) and (\ref{unfolding}),
the unfolding function in the bulk of the spectrum,
in the large $N$
limit is
\begin{equation}
\label{unf_a}
s(\epsilon) = \lambda\, {\rm sgn}(\epsilon)\,
\int_0^{|\epsilon|} d\epsilon' {1\over \epsilon'^{(1-\alpha)}} =
\lambda { |\epsilon|^{\alpha}}\,{\rm sgn} (\epsilon)
\;, \ \ \ \
 \lambda = {A\over 2\pi}\tan\left({\alpha \pi\over 2}\right)\; ,
\end{equation}
or
\begin{equation}
\epsilon(s) = \lambda^{-1/\alpha} |s|^{1/\alpha}\,{\rm sgn} (s)
\end{equation}
In order to simplify the derivation, we choose $\alpha = {1\over 2k+1}\, ,
k= 1,2,\dots $. In this case
\begin{equation}
\epsilon(s) = \left(s\over \lambda\right)^{(2k+1)}\; .
\end{equation}
The integral equation Eq.~(\ref{kern1}) is now
\begin{equation}
\label{kern1_a}
\int_{\infty}^{\infty}ds\, \psi(s')\ln|s^{(2k+1)} - s'^{(2k+1)}| = \varphi(s)
 +
{\rm const}\; .
\end{equation}
We change variables $s^{(2k+1)} = x\, , s_x = x^{1\over 2k+1}$ and we define
\begin{equation}
\tilde {\psi}(x) = {1\over 2k+1}{\psi}\, (s_x)x^{2k\over 2k+1}\; ,
\end{equation}
\begin{equation}
\tilde {\varphi}(x) = \varphi(s_x)\; ,
\end{equation}
\begin{equation}
\tilde{R_2}(x,x') = {1\over (2k+1)^2} (x x')^{-2k\over 2k+1}\,
R_2(s_x, s_{x'})\; ,
\end{equation}
in terms of which Eqs.\ (\ref{kern1}) and (\ref{kern2}) become
\begin{equation}
\int_{-\infty}^{\infty}dx'\, \tilde {\psi}(x')\ln|x-x'| =
\tilde {\varphi}(x)\; ,
\end{equation}
\begin{equation}
\tilde {\psi}(x) = \int_{-\infty}^{\infty}dx'\, \beta \tilde{R_2}(x,x')\; ,
\tilde {\varphi}(x')
\end{equation}
and the condition on $\tilde {\psi}(x)$ is now $\int_{-\infty}^{\infty}dx\,
\tilde {\psi}(x)=0$.
This is exactly the familiar case of pure logarithmic interaction
with strong confinement considered above
[see Eq.~(\ref{gauss_r2eq})]. The solution for the solving kernel
$\tilde{R_2}(x,x')$ is the WD universal correlation
given in Eq.~(\ref{gauss_r2}).
Using the relation between $\tilde{R_2}(x,x')$ and
${R_2}(x,x')$  and going back to the original variables,
$s$ and $s'$, we obtain
\begin{equation}
\label{r2_alpha}
{R_2}(s,s') = {(2k+1)^2\over \pi^2 \beta}\; .
{(ss')^{2k}\over (s^{2k+1}-s'^{2k+1})^2}
\end{equation}
This correlation function is not translational invariant. However
we restrict ourselves to the case in which both $s$ and $s'$ are
in the bulk and $\Delta s= |s-s'|<<s$. In this limit we can expand
the previous
expression in power of $\Delta s/s$. The correlation becomes
translationally invariant but what is even more important is that
we get back exactly Eq.~(\ref{gauss_r2}), namely
the universal WD correlation function. Thus we conclude, that at the
MFT level, the power-law ensemble has completely classical statistics, even
when the translational invariance is
broken in the thermodynamic limit and the unfolding procedure is
necessary.

We now consider the case of the double logarithmic potential.
The unfolding function in the bulk of the spectrum, $|\epsilon|>>1$ is now
\begin{equation}
s(\epsilon) = {A\over2}\, {\rm sgn}(\epsilon)\,
\int_0^{|\epsilon|} d\epsilon' {1\over \epsilon'} =
{A\over2} \ln|\epsilon|\,{\rm sgn} (\epsilon)\; ,
\end{equation}
or
\begin{equation}
\epsilon(s) = e^{2|s|/A}{\rm sgn}(s)
\approx 2\sinh \left({2s\over A}\right)\; .
\end{equation}

We plug this function into Eq.~(\ref{kern1}) and we perform the change of
variable, $\sinh\left({2s\over A}\right) = x$, with the corresponding
redefinition of $\psi$, $\varphi$ and $R_2$. Following the same procedure
used for the power-law potential, we arrive at
%
\begin{equation}
\label{r2_dblog}
R_2(s,s') = -{1\over \pi^2 A^2\beta}\,
{\cosh\left({2s\over A}\right)\cosh\left({2s'\over A}\right)
\over \sinh ^2\left({s-s'\over A}\right)\cosh ^2\left({s+s'\over A}\right)}
\; .
\end{equation}
Again the correlation function is not translational invariant. If $s$ and
$s'$ are both in the bulk, $|s'|\, , |s|>>1 $  and have the same sign,
we obtain the translational invariant expression
\begin{equation}
\label{normalR}
R_{2({\rm n})}(s,s') =- {1\over \pi^2 A^2\beta}\,
{1\over \sinh ^2\left({s-s'\over A}\right)}\; ,
\end{equation}
where the subscript $(n)$ stands for ``normal'' part.
If $|s-s'|>A$, the argument of $\sinh$ cannot be expanded and $A$
does not scale away.
Thus the correlations for the double
logarithmic potential are no longer universal and in contrast with
the power-law behavior of the WD class, they decrease
exponentially in agreement with the exact solution for $\beta =2$ by
Muttalib {\it et al.}.\cite{muttalib}

However there is one more great surprise, pointed
out in Ref.~[\onlinecite{RMpap2}] for $\beta=2$,
which is the re-appearance of
strong correlations at $s' \approx -s$. In fact when $s$ and $s'$ are in the
bulk but have different sign, $s' = -s + \Delta s$ we obtain
\begin{equation}
\label{anomalR}
R_{2(a)}(s,s') = -{1\over \pi^2 A^2\beta}\,
{1\over \cosh^2\left({s+s'\over A}\right)}\; .
\end{equation}
where the subscript $(a)$ stands for ``anomalous'' part.
This anomalous part of the correlation function breaks dramatically
the translational invariance. Its remarkable property is a narrow
correlation hole at $s'\approx -s$ with a depth, controlled by $A$,
that does not decrease
when $|s-s'| \simeq 2|s| \to \infty$ . Notice also that
the two regions where $R_{2(n)}(s,s')$ and $R_{2(a)}(s,s')$ are
nonzero are separated by a very large distance
when $s$ and $s'$ are in the bulk of the spectrum
and $N$ is large.

In Ref.~[\onlinecite{RMpap2}] a simplified application of
the method of the orthogonal polynomials, valid for $q= e^{-\pi^2 A} <<1$,
was used to derive
the two-level cluster function for the
case $\beta=2$, yielding the result:
\begin{equation}
\label{y2_dblog_pol}
Y_2(s,s') = {1\over \pi^2 A^2}\,
[\sin \pi(s-s')]^2\,
{\cosh\left({2s\over A}\right)\cosh\left({2s'\over A}\right)
\over \sinh ^2\left({s-s'\over A}\right)\cosh ^2
\left({s+s'\over A}\right)}\; .
\end{equation}
The normal part of $Y_2(s,s')$
\begin{equation}
\label{y2normdblogpol}
Y_{2({\rm n})}(s,s') = {1\over \pi^2 A^2}\,
\left [{\sin \pi(s-s')\,
\over \sinh \left({s-s'\over A}\right)}\right]^2\; ,
\end{equation}
is identical to the exact solution of Muttalib {\it et al.}\cite{muttalib}
for the
same small values of $q$ and it also identical to the exact
solution of Moshe {\it et al.}\cite{moshe} for the case of a RME with a
symmetry breaking term.

Here, using the MFT theory, we have generalized this result to any $\beta$.
As in all other MFT calculations, only the asymptotic form of the
$R_2$ is obtainable within the MFT treatment. In particular the
oscillatory function which vanishes as $|s-s'|\to 0$,
and thus gives rise to a residual level repulsion,
is totally out of reach. In its place we have $1/\beta$ which is the
average of the oscillations. Thus MFT and the method of orthogonal
polynomilas (which is exact) are completely consistent even in the case
of weak confinement where deviations from WD occur.

The appearance of anomalous correlations at $s'\approx -s$
is the result
of the system trying to develop Poissonian-like correlations
at $s'\approx s$,
while, at the same time, complying with the $U(N)$ invariance which
forces a normalization sum rule on $R_2$ to be satisfied even in the
large $N$ limit.\cite{RMpap2} The sum rule reads
\begin{equation}
\label{sumR}
\int_{-\infty}^{\infty} R_2(s,s')ds' =0\; ,
\end{equation}
or, in terms of $Y_2(s,s')$,
\begin{equation}
\int_{-\infty}^{\infty} Y_2(s,s')ds' =1\; ,
\end{equation}
and it is must be satisfied in the case of a $U(N)$ invariant RME,
because of the long-range
nature of the universal logarithmic interaction always present
in such ensembles.\cite{RMpap2} The normal part alone of $Y_2(s,s')$, given in
Eq.~(\ref{y2normdblogpol}), does not satisfy Eq.~(\ref{sumR}).
But the sum rule deficiency
\begin{equation}
\label{etaD}
\eta = 1 - \int_{\infty}^{\infty} Y_{2({\rm n})}(s)\; ds\; ,
\end{equation}
is taken care of by the anomalous
correlations\cite{noteSUMR}:
\begin{equation}
\label{etaDAn}
\int_{\infty}^{\infty} Y_{2({\rm a})}(s)\; ds = \eta\; .
\end{equation}
However, no matter how this is
realized in practice, what is important is firstly the fact that the
$U(N)$ invariant RME with soft confinement manages to develop
the exponentially decaying two-level correlation function of
Eq.~(\ref{y2normdblogpol})
in an infinitely
large energy region in the bulk of the spectrum, where the anomalous
correlations are irrelevant. Secondly, the normal part of correlation function
is exactly equal to the
expression obtained for the RME (\ref{ShE}), where the symmetry is explicitly
broken.
This occurrence has been interpreted in Ref.~[\onlinecite{RMpap2}]
as a signal of the
spontaneous
breakdown of the $U(N)$ invariance in case of soft confinement,
with the parameter $\eta$ playing the role of the order parameter.

The physical interpretation of the anomalous correlations at
$s'= \approx -s$ in terms of Coulomb plasma
is very simple if one looks at how the logarithmic
pair-wise interaction is transformed in the unfolded variables.
\begin{equation}
f(s,s') = - \ln|\sinh(2 s/A) - \sinh(2 s/A)|\; .
\end{equation}
In the unfolded variables the interaction is no longer translational
invariant. We can try to rewrite it in a form that looks more
translationally invariant in the following way:
\begin{eqnarray}
f(s,s')&=&-\ln\left|2\sinh\left({s-s'\over A}\right)
\cosh\left({s+s'\over A}\right)\right|
\nonumber\\
&=& -\ln\left| 2\sinh\left({s-s'\over A}\right)\right| -
  \ln\cosh\left[{s'-(-s)\over A}\right]\; .
\end{eqnarray}

We see that the interaction splits into two terms: the first is an ordinary,
translationally invariant, repulsion between two particles located at
$s'$ and $s$; the second, however, represents
the interaction between a particle at position $s'$ with the {\it image charge}
at $-s$ of a particle at position $s$. In other words, a particle at position
$s$ will repel particles aroung it. But its image with respect to the
origin will also repel
particles around the the position $-s$. Notice that the image charge term
depends on $A$ -- it increases when $A$ decreases-- but it is less singular
that the direct term because of the functional dependence on $\cosh(s+s')$.

So far we have always restricted ourselves to bulk properties of the
correlations. As we will show in more detail in the next section, there
are some interesting effects in the correlation function in the center
of the spectrum. There, even the power-law potential for $\alpha<1$
displays deviations from the universal WD behavior.

\section{Monte Carlo Simulations}
\label{sec5}

The statistical properties of the onedimensional classical system\cite{note1}
in thermodynamical
equilibrium, whose probability distribution is given by
Eqs.~(\ref{boltz})-(\ref{h}),
can be conveniently studied by carrying out Monte Carlo (MC) simulations.

The MC method is useful because it is not restricted to a
particular value of $\beta$, as for the method of orthogonal polynomials,
and it is very accurate in all the regions of the spectrum,
in contrast with MFT which is good only in the bulk.
It also allows the evaluation of important statistical quantities like the
level spacing distribution function (LSDF) and the number variance in
a straightforward way. Finally it can be used to study level statistics where
the ``particle'' interaction is more general than the simple
logarithmic interaction considered in RMT.
These more complicated interactions
have been shown to play an important role in some disordered
systems.\cite{beenakker93prl}

The nature of MC simulations is best illustrated with an example.
Suppose we want to calculate the mean particle density $\rho(\epsilon)$ as
a function of the position $\epsilon$. According to Eq.~(\ref{mean_density})
we need to perform a multi-dimensional integral, which up to an overall
normalization, is equivalent to a `thermal' average over an ensemble
of particle configurations. The MC method replaces this ensemble average
with a `time' average, but the time evolution is determined by equations
that are artificial and chosen for convenience. To calculate $\rho(\epsilon)$
one partitions the real axis into bins with boundaries $\epsilon_n$
determined by
\begin{equation}
\epsilon_n= n\Delta \epsilon\; , \ \ \ \ \ n=\pm 1,2,3,\dots
\end{equation}
where $\Delta \epsilon$ is the with of the bins. At the end of each
time step (to be defined below) we obtain an updated configuration
$\{\epsilon_i\}\, , i=1,\dots , N$, and we add one to a bin if a particle
in this configuration lies in it. As `time' progresses,
the number of particles
in a bin will become proportional to the mean density density at the position
where the bin is centered. For the evolution density we have taken a simple
Metropolis algorithm, which works in the following way. At each time
step or sweep, we scan through the particles and attempt to move each one.
Actually in each sweep we pick $N$ times one particle at random
in the system, so it is
possible that in one particular sweep one particle is chosen more than once
and another is not touched. The moving attempt involves picking at random
any position between the particle the proceeds and the
one that follows
the particle that we are trying to move and taking this position as the new
attempted position. The attempted move is chosen in this particular way
simply to optimize the convergence rate of the algorithm. It has the
important property that if we start with an order sequence of particles,
$\epsilon_1<\epsilon_2 < \dots \epsilon_N$, the sequence remains ordered
in the time evolution. To decide whether or not to accept the move,
we calculate the change $\Delta E$ to the system's energy that would occur
if the particle were moved to the new position. If $\Delta E$ is negative
the move is accepted, and the particle is given the new position. If
$\Delta E$ is positive the move is accepted conditionally. One picks
a random number between $0$ and $1$ and accepts the move if this number
is smaller than $\exp(-\beta \Delta E)$. Before measuring any quantity
the system must reach equilibrium and this is obtained by a certain
number of ``warming-up'' sweeps that reach a `typically'
sampled configuration.

We have carried out simulations over systems with up to $200$ particles.
We noticed that the simulations are in all cases very
stable even for smaller $N$, and therefore we have typically worked with
systems of $N=100$ particles. Equilibration is usually reached very
fastly and we have typically used $10^5$ sweeps to warm up the system.
The averaged are taken over $10^6$ sweeps and the statistics that we
are able to obtain are usually excellent.

To make sure that the method
works and is able to give numerically accurate results we have first
studied the three Gaussian Ensembles whose density, two-particle correlation
function, spacing distribution and particle variance are exactly
known.\cite{mehta}. For these ensembles the MC simulations reproduce
the known result very accurately. For example, in the
calculation of $P(\sigma)$ the method is able
to detect the small deviation that there is between the exact result and the
Wigner surmise.

We now proceed to discuss in detail the results for the different quantities
of interest for the case of power-law and double logarithmic ensemble.

\subsection{The density of states}

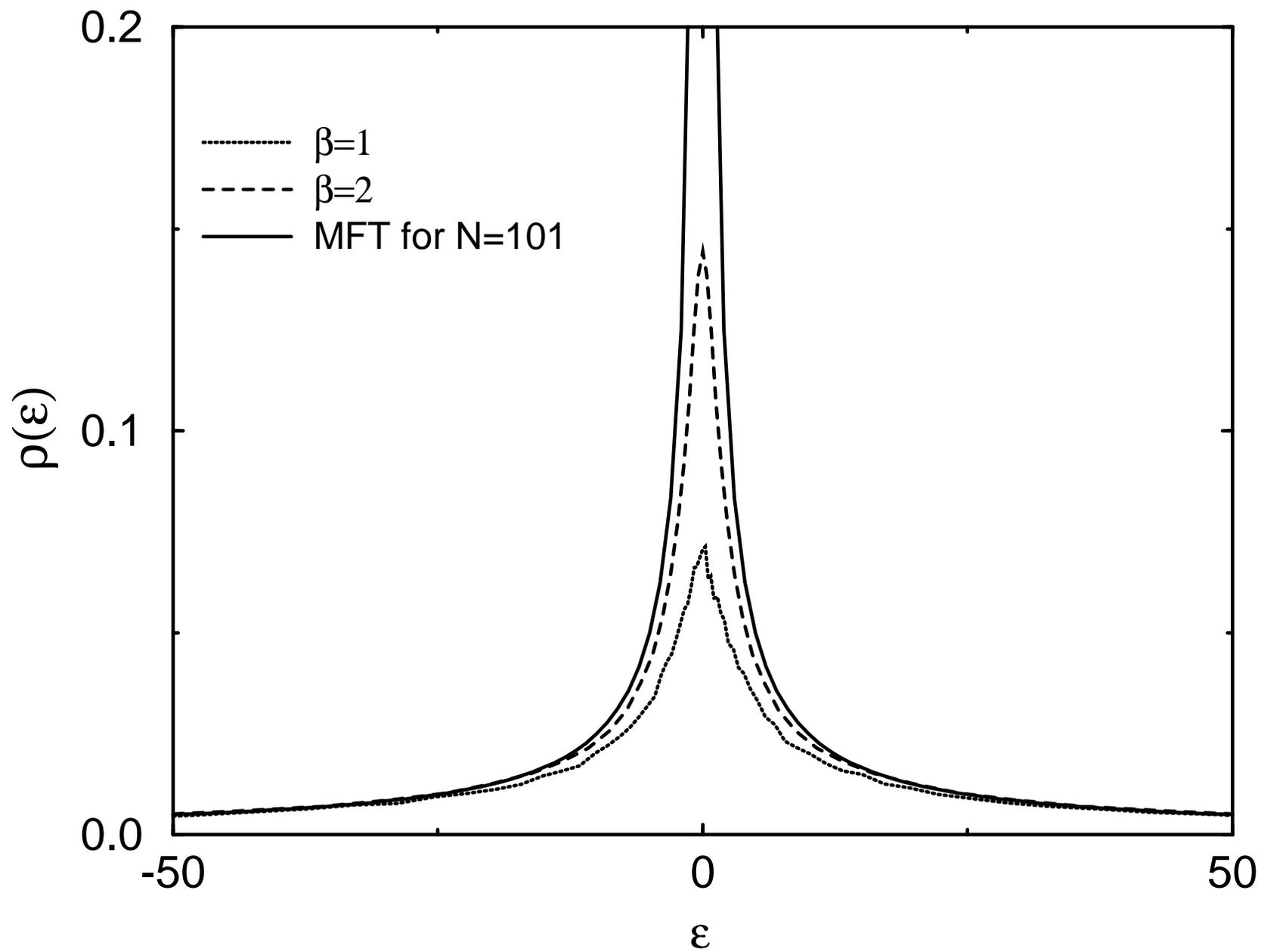
\begin{figure}
\caption{Density of states for the logarithmic potential for $A=0.5$.
The Monte Carlo (MC) results for $\beta=1,2$ are plotted
in small region around the origin $\epsilon=0$ with the
MFT density, which corresponds to $\beta=\infty$ and diverges at $\epsilon=0$.
For $\beta=1,2$ and $4$(not shown), the density is finite at $\epsilon=0$.
All the curve rapidly collapse on
top of each other away from the origin.}
\label{fig1}
\end{figure}

The MC evaluation of the average density of states $\rho(\epsilon)$
is carried out as explained in the example above. In Fig.~\ref{fig1}
we plot this quantity for the logarithmic potential for $A=0.5$
and $\beta=1,2$. The case of weak power-law potential, with
$0\le \alpha <1$ is qualitatively the same\cite{RMpap1}.
The agreement between the MC result and the MF
expression $\rho_{\rm MF}$ given in Eq.~(\ref{log2p}) is very good
except around the origin, where in contrast to MFT, the simulations
give a sharply picked but finite density at $\epsilon=0$. For $\beta=2$
the MC result coincides with the value obtained by the method of
orthogonal polynomials.\cite{kravtsov_ump}
The first hundred of these for the power-law potential
can be easily generated numerically\cite{kravtsov_ump}
and the density at $\epsilon=0$
obtained from them converges very fast. This fast N independence of the
center of the spectrum is also seen very well with the simulations and
it is an important property of the particle density for weak confinement.
We refer to it as the ``incompressibility'' of the core of
the particle-density
distribution. In contrast to the case of strong confinement,
$V(\epsilon) \sim |\epsilon|^{\alpha}, \alpha \geq 1$,
where the density at the center of the spectrum scales with $N^{\alpha -1}$,
for
$\alpha <1$ the confining potential is too weak to ``compress'' the
particle in the core region near the origin.
After the initial formation of the sharp but finite
peak at $\epsilon=0$ (which happens for $N<<100$),
on adding more
particles to the system, these always go to the ends of the distribution
instead of spreading homogeneously throughout the spectrum. The particle
density in the core of the spectrum is almost independent of $N$ but
depends on the inverse temperature $\beta$.
In the bulk of the spectrum the density decreases like $1/\epsilon^{1-\alpha}$
in the $N\to \infty$ limit. Therefore translational invariance is broken in the
thermodynamic limit, in agreement with the MFT.

\subsection{The two-level correlation function}
The MC evaluation of the two-point correlation function $R_2(s,s')$, as for
any other
correlation function, faces the complication of the breakdown of translational
invariance. Therefore we need to carry out a numerical unfolding of the
spectrum in order to compare with the classical statistics. We have
considered three different unfolding procedures that can be used in
different circumstances. In the simplest case we are interested in the
{\it bulk} correlations. Therefore the simplest unfolding scheme
consists in carrying out, for each MC configuration generated at ``time'' $t$,
$\{\epsilon_i\}^t\;, i=1,\dots\;, N$, the mapping
\begin{equation}
\{\epsilon_i\}^t \to \{s_i\}^t\;, \ \ \ \ \
s_i={\rm sgn}(\epsilon)\int_0^{|\epsilon|}d\epsilon\;
\rho(\epsilon)_{\rm MF}\; .
\end{equation}
The unfolded configurations $\{s_i\}^t$ generated in this way are then used
to measure the correlations, which will be automatically in the right units.

The second method, which turns out to be particularly useful for
the logarithmic potential, consists in performing the change of variable
$\epsilon \to s = s(\epsilon)$ directly in the joint probability
density function
\begin{equation}
{\cal P}(\{ \epsilon_i\} ) \to {\tilde {\cal P} }(\{ s_i\}) \propto
\exp [-\beta{\tilde {\cal H}}(\{s_i\})]\; ,
\end{equation}
\begin{equation}
{\tilde {\cal H}}(\{s_i\}) = - \sum_{i,j}\ln|\epsilon_{s_i} - \epsilon_{s_j}|
+ \sum_i V(\epsilon_{s_i})
-{1\over \beta}\sum_{i}\exp[\ln (1/\rho(\epsilon_{s_i})]\; .
\end{equation}
Notice that the one-body confining potential is modified by a
$\beta$-dependent term
coming from the Jacobian of the transformation.

Both these two unfolding schemes make use of the MFT particle density
and therefore can only be used to study properties in the bulk. In order
to study the correlation function around the origin we need a more
precise expression for $\rho(\epsilon)$. Therefore we carry out the
following procedure.
The unfolded two-level correlation
function at $s=0$ can be expressed as
\begin{equation}
R_2(0,s') = R_2\left (0,\epsilon_{s'}\right)\; .
\end{equation}
The function $\epsilon(s)$ is now obtained by numerically inverting
$s(\epsilon)= \int_0^{\epsilon} \langle \rho(\epsilon ')\rangle \;
d\epsilon '$,
with the density $\langle \rho(\epsilon)\rangle$ evaluated directly by the
MC simulations.

Once we have taken care of the unfolding,
the evaluation the two-level correlation function $R_2(s,s')$
by MC is very simple:
we fix a reference particle at $s$
and then we compute  the ``conditional'' particle density
$\rho(s')|_s$
of all the remaining particles with respect to the reference point,
using a partition in bins as explained before.\cite{noteR2}
In studying bulk properties the reference particle can
be let free to move, since in the unfolded coordinates the
density is constant. In this case $s$ in $R_2(s,s')$ must be
interpreted as $\langle s\rangle$.
Because of the particular way of choosing the trial moves adopted here, the
standard deviation
from $\langle s\rangle$ turns out to be small once the system has reached
equilibrium.

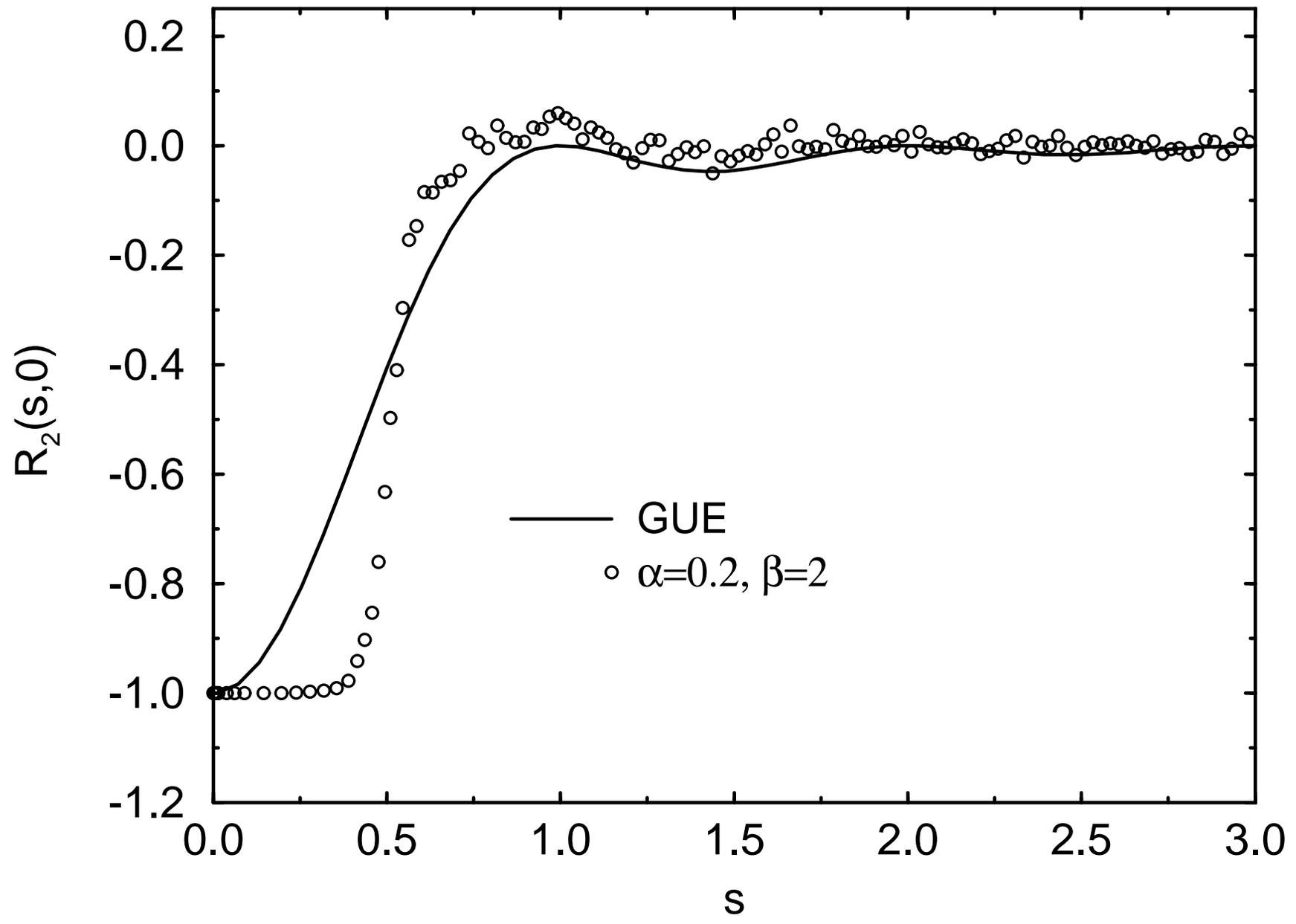
\begin{figure}
\caption{The MC two-level correlation function $R_2(s'=0,s)$ vs. $s$ for the
power-law potential with $\alpha=0.2$, $\beta=2$. The solid line is the
result of the GUE, $R_2(r) = -\left[{\sin(\pi r) \over \pi r}\right]^2$.
[We have omitted the $\delta$ function at the origin].}
\label{fig2}
\end{figure}

For the power-law potential the Monte Carlo simulations show
that the two-level correlation function $R_2(s,s')$ is perfectly equal to
the WD expression for any $\alpha$ when the reference particle is in the bulk
of the spectrum, in agreement with the MFT and orthogonal polynomial
results.
However this universality is broken around the origin. In Fig.~\ref{fig2} we
plot $R_2(0,s)$ for  $\alpha =0.2$, and $\beta=2$ (the $\delta$ function
is not included).
For small $s$,  $R_2(0,s)$ does not follow the classical universal
behavior $s^{\beta}$ but rather starts out like $s^{\beta/\alpha}$.
Thus we have a sort of ``super Wigner'' behavior, with stronger level
repulsion at short separation the smaller is $\alpha$.
Since the cluster function must satisfy the normalization sum rule,
this implies a faster decay of $R_2(0,s)$ at large $s$. Using
the method of orthogonal polynomials one can show that this
decay goes like $1/s^{(1+ 1/\alpha)}$.\cite{kravtsov_ump}

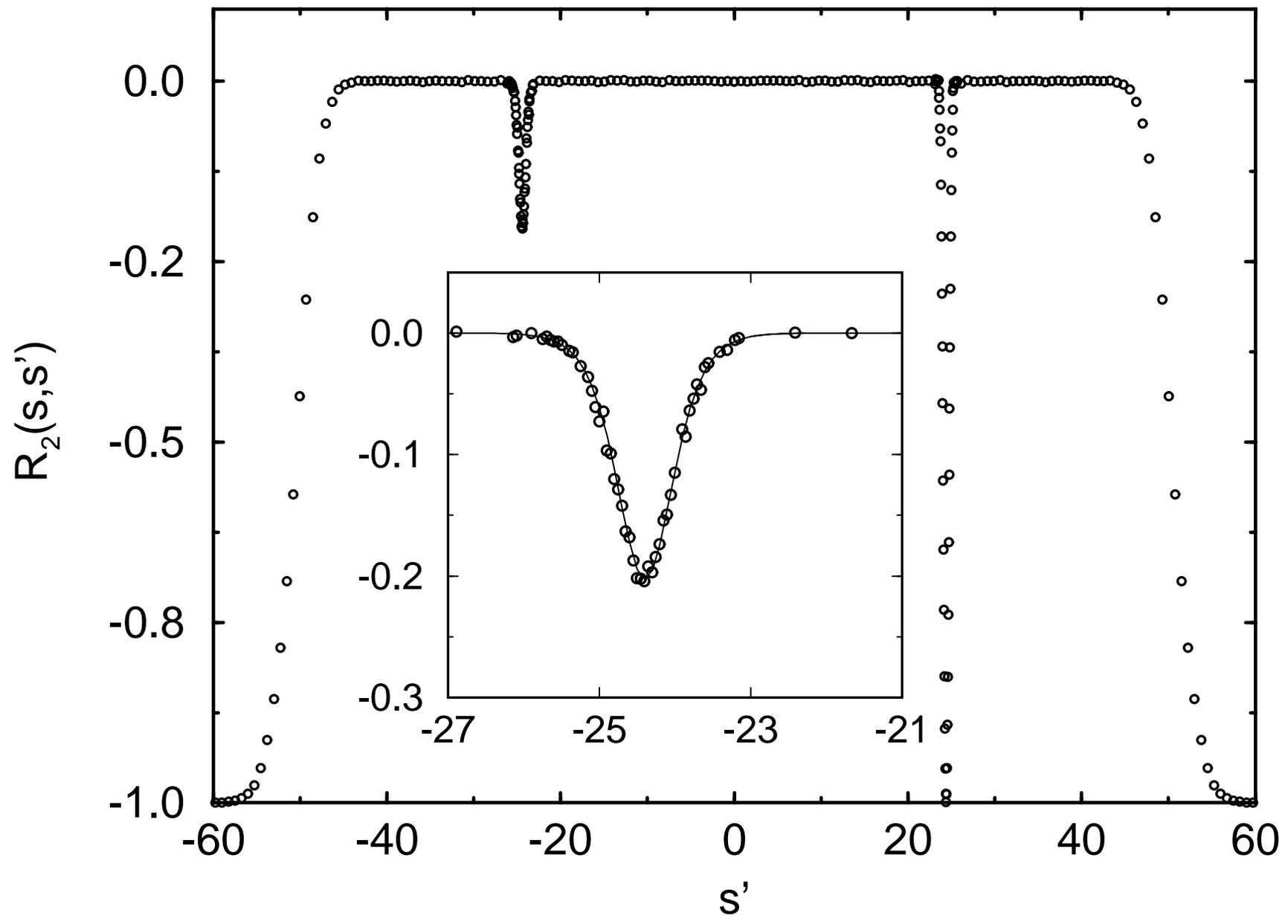
\begin{figure}
\caption{MC result for the two-level correlation function for the
logarithmic confinement, Eq.~(\protect\ref{Vdbl}),
showing the existence of a ``ghost'' hole at $s'=-s$.
The simulations are performed for $\beta=2$ and $A=0.5$ with $N=101$
particles. The reference
particle is mobile around $s\approx 24.4$. The solid line in the inset
corresponds to Eq.~(\protect\ref{anomalR}).}
\label{fig3}
\end{figure}

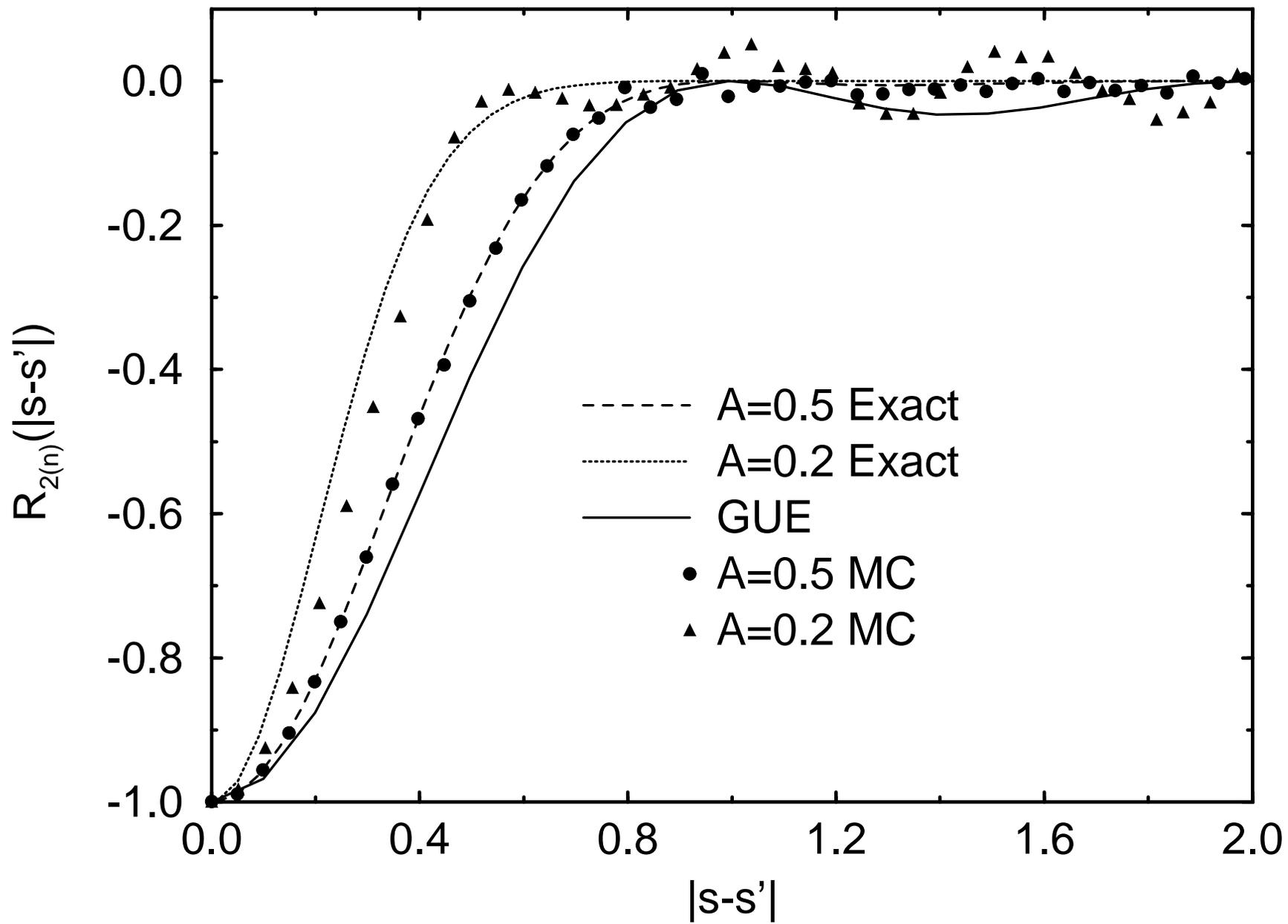
\begin{figure}
\caption{MC results for the normal part of $R_{2(n)}(s,s')$
$\beta =2$, plotted together with the
corresponding exact expressions, Eq.~(\protect\ref{normalR}),
and with the GUE
curve as a comparison. Notice the good agreement at small $s'$
between numerical
and exact results for $A=0.5$ which becomes worse for $A\le 0.2$.
The fluctuations at $s'>1$, more visible for $A=0.2$, are due to
finite-size fluctuations of the exact density, which the
unfolding procedure via
the MFT density cannot cure. These fluctuations are much smaller
for $\beta=1$.}
\label{fig4}
\end{figure}

We now discuss the bulk properties of
$R_2(s,s')$ for the logarithmic potential.
As shown if Fig.~\ref{fig3}, the MC simulations confirms fully the
surprising result of the bulk breakdown of the translational invariance in
$R_2(s,s')$ and the appearance of the ``ghost'' correlation hole at $s=-s'$.
The reference particle was let free to move in the positive part of
the spectrum around $\langle s \rangle = 24$ (in unfolded coordinates).
Besides the usual correlation hole around this position, another
one appears symmetrically with respect to $s=0$, as if there was
an image of the reference particle, the ``ghost'', located around
$\langle s \rangle = -24$. The contribution of the anomalous part
increases upon decreasing $A$. For values of the parameter $A$ not
too small, both parts of the cluster functions obtained
numerically
are in good agreement with the analytical results given
in Eqs.\ (\ref{normalR})
and (\ref{anomalR}), as seen in Fig.~\ref{fig4} where the normal
part of $R_2$ is plotted. However, as one can also see
from the same figure, already
for $A=0.2$ the MC result for the normal part starts to deviate
from the analytical
formula, which in fact becomes invalid for $A< 1/\pi^2 \approx 0.1$.
In this regime the MC simulations show that the cluster function instead
of decaying exponentially at very small $s$, starts out more and more
flat and in the limit of very small $A$ it converges to a box or ``well'' of
width 1/2 and depth (-1). A similar behavior occurs for the anomalous
part as well. Therefore, in the $A\to 0$ limit,
the cluster function is composed
of two rectangular wells centered at $s$ and $-s$. This result can also
be obtained from the exact solution by Muttalib {\it et al.},\cite{muttalib}
plotting
their general expression for $R_2$, which remains valid in the
regime of very small $A$.

We conclude that the correlation function $R(s,s')$
of the double logarithmic ensemble,
displays a cross-over from WD toward a Poissonic behavior for
{\it intermediate} $A$, i.e. $1/\pi^2< A<1 $, but never really
becomes exactly
equal to $\delta(s-s')$ of the Poisson distribution.

\subsection{The Level Spacing Distribution Function (LSDF)}

Let us consider a sequence of successive levels
$\epsilon_1 \leq \epsilon_2 \leq \dots$ and let $S_1,S_2,\dots$ be
their distance apart, $S_i = \epsilon_{i+1} -\epsilon_i$. The average
value of $S_i$ is the mean level spacing $\Delta$. We suppose for
the moment that the average density and therefore
$\Delta= \rho^{-1} = const$.
We further define the relative spacings $\sigma_{i}=S_i/\Delta$.
The nearest neighbor LSDF,
$P(\sigma)$, is defined by the condition that $P(\sigma)ds$ is the
probability
that any $\sigma_{i}$ will have a value between $\sigma$ and $\sigma+d\sigma$.

If the energy levels are completely uncorrelated, one can immediately
prove that the LSDF is the Poisson distribution,
\begin{equation}
\label{p_p}
P_p(\sigma) = \exp(-\sigma)\; .
\end{equation}
In contrast, for a large class of chaotic or disorder systems where
the energy levels are correlated, $P(\sigma)$ is very well described
by the so called Wigner surmise,\cite{note2}
\begin{equation}
\label{p_w}
P_w(\sigma) = {\pi \sigma\over 2} \exp(-{\pi\over 4} \sigma^2)\; .
\end{equation}

The Wigner surmise vanishes at short separations, showing the phenomenon
of level repulsion, typical, for example, for extended wave functions
of a disordered conductor. The Poisson distribution, in contrast, allows
level degeneracy, as in the case of an Anderson insulator, where the
wave functions are localized and do not overlap. Notice that $P_w(\sigma)$
falls down faster  at large $\sigma$ than $P_p(\sigma)$.
This is again due to level repulsion
which, in a finite energy window, prevents the appearance of large
energy gaps with no levels in them.

For the Gaussian Ensembles one can derive exact expressions for $P(\sigma)$,
which turn out to be very close but not identical to the
Wigner surmise.\cite{mehta}
All the RME with logarithmic interaction and strong confinement
belong to the GE (or WD) universality
class and thus their $P(\sigma)$ is also very close to the Wigner surmise.

The analytical determination the LSDF is not straightforward.
For $\beta =2$ the function $P(\sigma)$ can be expressed in
terms of a determinant of
the two-level cluster function\cite{mehta},
which in general must be evaluated numerically.\cite{muttalib}
Alternatively one can use MFT\cite{AKLplasma,KLplasma,eriksen}
but this method has not been
extended yet to the potential (\ref{Vdbl}).
On the other hand the LSDF is
easily calculated by MC.
In terms of
the plasma model, $P(\sigma)$ is defined, once the system has been unfolded,
as the probability density of finding the nearest adjacent particle
at a distance $s$ from a given reference particle.
The LSDF obviously coincides with the the two-level correlation for
very small $\sigma$.

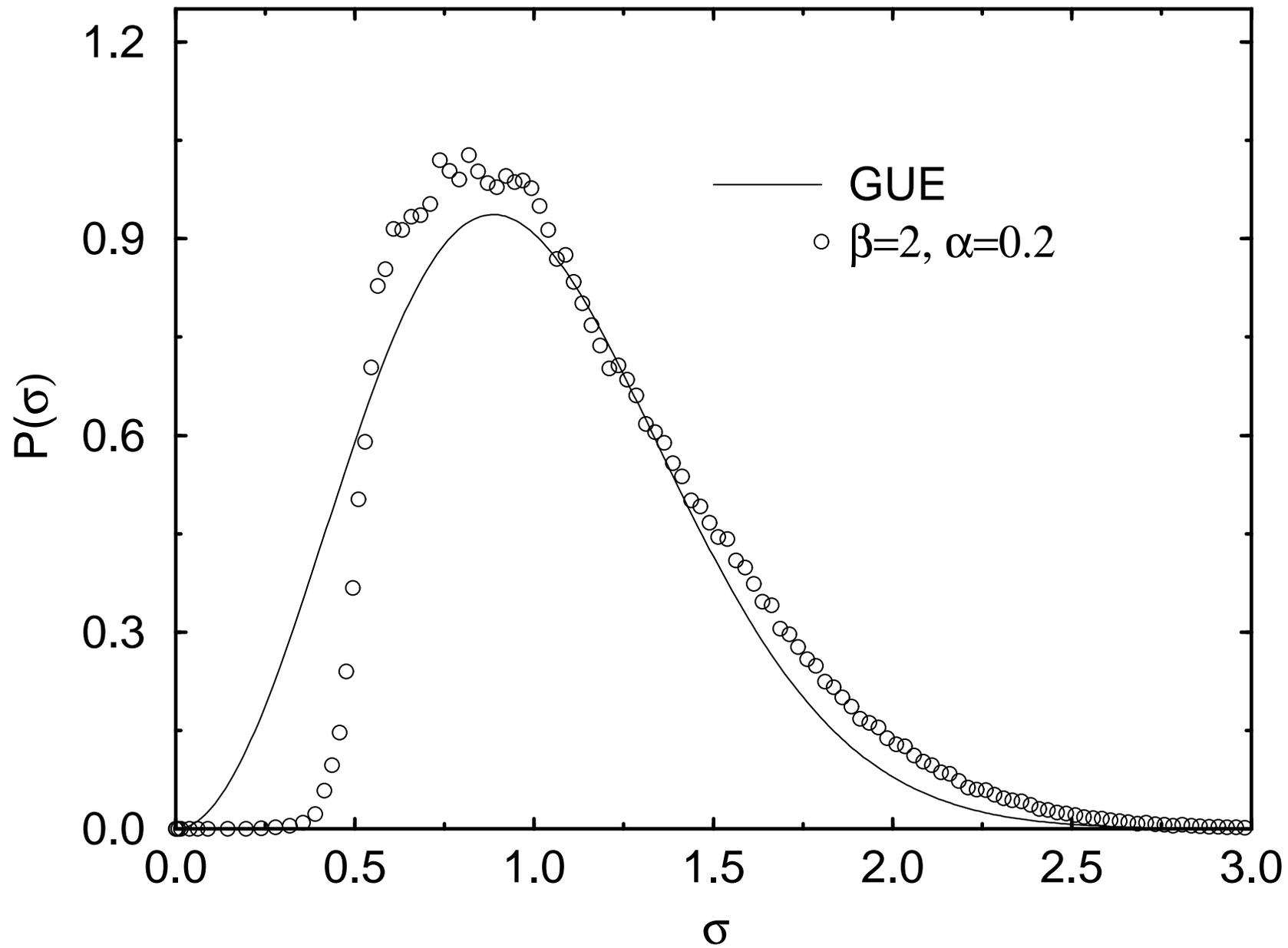
\begin{figure}
\caption{MC result for the LSDF in the middle of the spectrum
of the power-law potential
with $\alpha=0.2$ and $\beta=2$. For $\sigma \to 0$ $P(\sigma)$ vanishes
like $\sigma^{\beta/\alpha} =\sigma^{10}$.
The exact GUE distribution is also
plotted.}
\label{fig5}
\end{figure}

{}From the results of the two-level correlation function we
expect the LSDF for the power-law potential
to be identical to the Wigner surmise in the
bulk, with possible deviations at the origin.
The MC simulations confirm fully these expectations.
To calculate the unfolded spacing around the origin, we
fix a particle at $\epsilon=0$ and we perform the
unfolding by computing
\begin{equation}
P(\sigma) = \left [ P(\epsilon)\over \langle \rho(\epsilon)\rangle\right ]_
{\epsilon=\epsilon(\sigma)}\; ,
\end{equation}
where the function $\epsilon(\sigma)$ is again obtained by numerically
inverting $\sigma(\epsilon) =
\int_0^{\epsilon} \langle \rho(\epsilon ')\rangle \; d\epsilon '$,
The result is shown in Fig.~\ref{fig5}, where we plot the
LSDF for $\alpha 0.2$ and $\beta=2$. The classical
spacing for the Gaussian Orthogonal Ensemble ($\alpha =2$) is also plotted.
The figures clearly shows the deviations of $P(\sigma)$ from the GOE result.
In particular for small $\sigma$ the LSDF does not follow the universal
behavior $\sigma ^{\beta}$ of the Wigner surmise but starts out like
$\sigma^{\beta/\alpha}$. This is  the same ``super Wigner'' behavior
already found for the two-point correlation function.

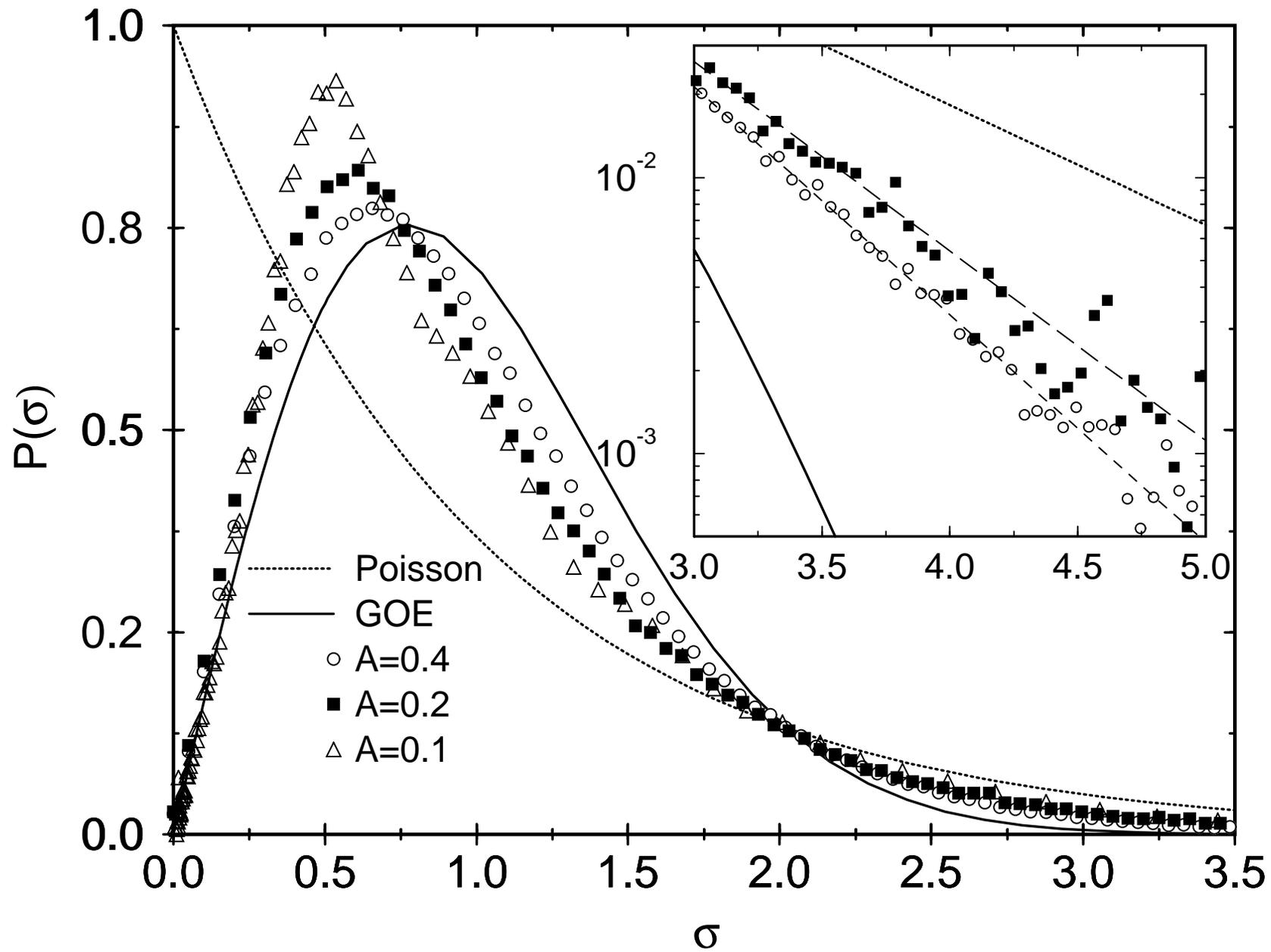
\begin{figure}
\caption{MC results for the bulk LSDF of
the logarithmic potential
with $\beta=1$ and different values of $A$ showing a cross-over betweem
and the GOE and the Poisson distributions,
also plotted. For $A<0.2$ the LSDF stops approaching the Poisson
function and tends to a single $\delta$ function peak at $\sigma\approx 1$.
Shown in the inset is the large $\sigma$ behavior of $P(\sigma)$. Notice the
logarithmic scale for the $y$-axis. The two dashed straight lines are
fitting functions of the form $exp(-a(A,\beta)\sigma)$ }.
\label{fig6}
\end{figure}

In Fig.~\ref{fig6} we show the {\it bulk} LSDF for the double logarithmic
potential in the case of the Orthogonal Ensemble. We plot
$P(\sigma)$ for several values of the parameter $A$, together with
the distribution of the Gaussian Orthogonal Ensemble and Poisson
distribution for comparison.
We see that,
for those values of $A<1$ for which the two-particle correlation
function displayed a deviation from the classical GOE behavior, we
have a corresponding deviation form the classical LSDF
toward a more Poisson-like behavior. $P(\sigma)$ still starts out
linearly at small $\sigma$, but the initial slope increases upon
decreases $A$ as a result of a smaller level repulsion. The
peak of the distribution shifts from $\sigma \approx 1$ to smaller
values $\approx A$. For large separations the decay is also slower than the
the GOE result. In fact, as we show in the inset, plotting
$\ln P(\sigma)$ vs. $\sigma>>1$, we can fit rather well the curves with
straight lines
\begin{equation}
\ln P(\sigma) \sim -a(A,\beta)\sigma\; , \qquad \sigma>>1\; ,
\end{equation}
where the constant $a(A,\beta)>1$ decreases with increasing $A$.
Notice that all curve cross at the same point
at $\sigma \approx 2$. Similar features and deviations from the
Gaussian Ensemble are obtained also from the other two symmetries,
unitary and symplectic.

The cross-over toward the Poisson distribution stops, however, at
around $A\approx 0.2$. We have shown that for $A<0.2$ the
correlation function, instead of becoming closer and closer to
a $\delta$ function when $A$ is further decreased, it
turns around and for very small
$A$ it approaches instead a
square well of width 1/2. Something similar happens for the LSDF.
We can already see in Fig.~\ref{fig6} that for
$A=0.1$ the initial slope of the distribution has stopped increasing,
and the height of the peak is getting close to one. For yet smaller
values of $A$ (not shown in the picture), the initial slope starts
{\it decreasing} and the LSDF,
instead of
approaching the Poisson distribution,
will tend eventually to a single narrow peak of height $>>1$, centered at
$\sigma \approx 1$.

\subsection{The Number Variance}

So far we have considered the correlation functions that probe
essentially the local fluctuations of a small number $n \simeq 1$ of energy
levels. We now turn to the variance
${\rm var}(\langle n\rangle) = \langle (n- \langle n\rangle)^2 \rangle =
\langle n^2 \rangle - \langle n\rangle ^2$ of the number of levels in
an energy window that contains $1\ll \langle n\rangle \ll N$ on the average.
The number variance is a statistical quantity
that provides a
quantitative measure of the long-range rigidity of the energy
spectrum.

For the Poisson distribution the levels are uncorrelated and there are large
level-number fluctuations, leading to a {\it linear} variance
\begin{equation}
{\rm var_p}(\langle n\rangle) = \langle n\rangle\; .
\end{equation}
On the other hand, the level correlations in the WD statistics makes
the spectrum more rigid and the number variance grows only {\it logarithmic}
\begin{equation}
\label{var_w}
{\rm var_w}(\langle n\rangle) = {2 \over \pi^2 \beta}\ln \langle n\rangle +
C_{\beta}\ + O(1/\langle n\rangle)\; ,
\end{equation}

where $C_{\beta}$ is a constant of order 1, which depends on the symmetry.

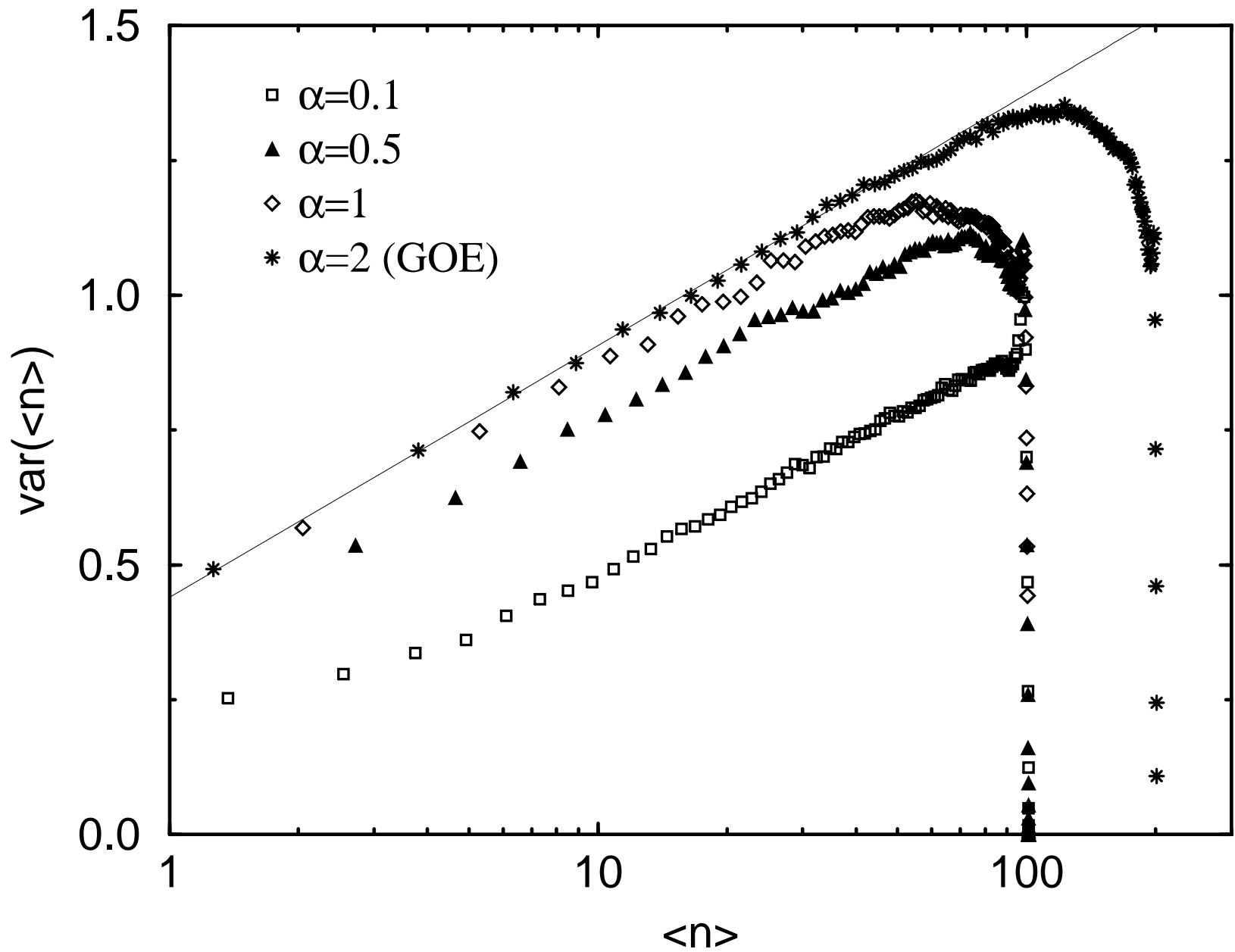
\begin{figure}
\caption{The MC level number variance vs $\langle n\rangle$
for the power-law
potential, for $\beta=1$ and several values of $\alpha \le 1$.
The dashed curve is the GOE result given in Eq.~(\protect\ref{var_w}).
The energy windows containing $\langle N \rangle$ particles are centered at
the origin $\epsilon=0$. The nonuniversal ``super-Wigner'' behavior
of the correlations
at $\epsilon=0$ is responsible of
the $\alpha$-dependence of the constant term in the variance, which is
otherwise equal to the that one of GOE.}
\label{fig7}
\end{figure}

The MC results for the power-law confinement show that Eq.~(\ref{var_w})
is perfectly satisfied, for every $\alpha$, in the bulk, namely when
the energy windows does
not contain the origin with its nonuniversal correlations.
On the other end when the energy windows are centered at the origin,
the ``super Wigner'' correlations present at $\epsilon=0$
for $\alpha<1$, manifest themselves
making the constant $C_{\beta}$ in the number variance $\alpha$ dependent.
As we show in Fig.~\ref{fig7}, for $\beta=1$, $C_{1}$
decreases with $\alpha$, when
$\alpha<1$, because there is more level repulsion in the area of the origin
and therefore more
level rigidity in the overall spectrum.
For $\alpha \geq 1$ the universal value of the Guassian Ensemble
$C_1 \approx 0.4420$,
is recovered.
The $\alpha$ dependence of $C_{\beta}$ is
however the only deviation from universality in the number variance,
the logarithmic
dependence being unchanged for the power-law confinement.

We now come to the case of the logarithmic potential.
We have seen that the presence of the ``ghost'' correlations break
translational invariance in the two-point correlation function.
It was shown in Ref.~ [\onlinecite{RMpap2}] that, due to such a
breakdown of translational
invariance, the number variance for the double logarithmic confinement
depends on the position of the energy window in a very essential way.
If the energy window does not contain the
origin, then the effect of the ghost peak is not felt and the system
is Poisson-like, with translationally invariant correlations
(in the energy range considered)
given by Eq.~(\ref{normalR}). In this case the number variance is
also Poisson-like and increases
{\it linearly}, as one expects in
presence of exponentially decaying correlations.
The coefficient of the linear term $\eta$ is less than 1,
and it is given by Eq.~(\ref{etaD}).
Here $\eta$ is nonzero because the normal part alone of the
correlation function
fails to satisfy the normalization sum rule. It increases upon decreasing $A$
because the ``spectral weight'' of the normal part of the cluster
function (\ref{y2normdblogpol}) decreases.
However,
if the energy windows are symmetric with respect to the origin, the
ghost correlations become effective and their contribution allow
the sum rule to be satisfied. Therefore the coefficient of
linear term $\eta$ in the variance vanishes in this case.

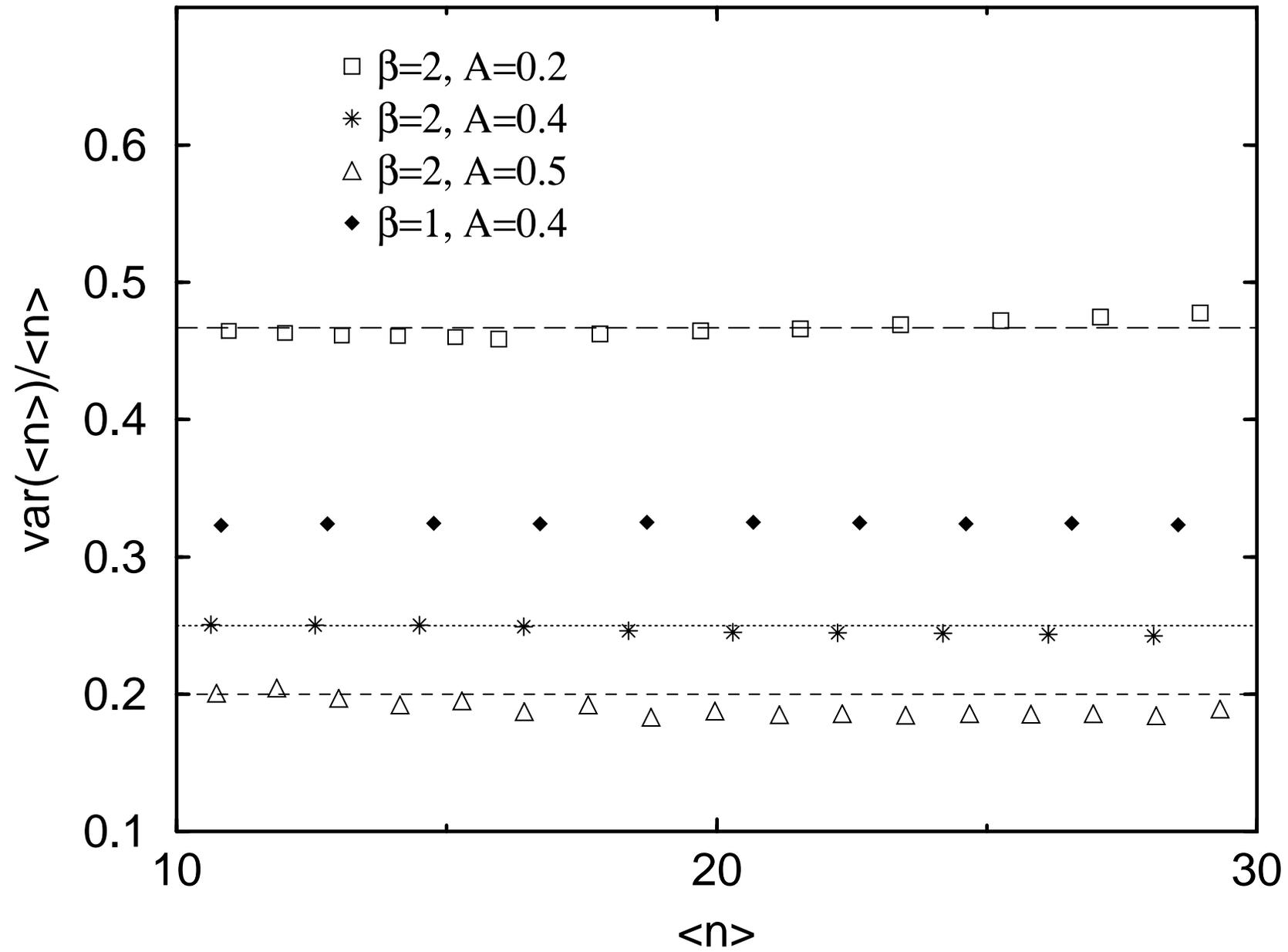
\begin{figure}
\caption{The relative variance
${\rm var}(\langle n \rangle/ \langle n\rangle$ vs $\langle n\rangle$
for the logarithmic potential
for different values $A$ and $\beta$. The energy window are centered in
the bulk of the spectrum and the variance is linear with $\langle n\rangle$.
The constant straight lines are the slopes of the linear term of the
exact solution of Ref.~[\protect\onlinecite{blecken}]
for $\beta=2$ and corresponding $A$.}
\label{fig8}
\end{figure}

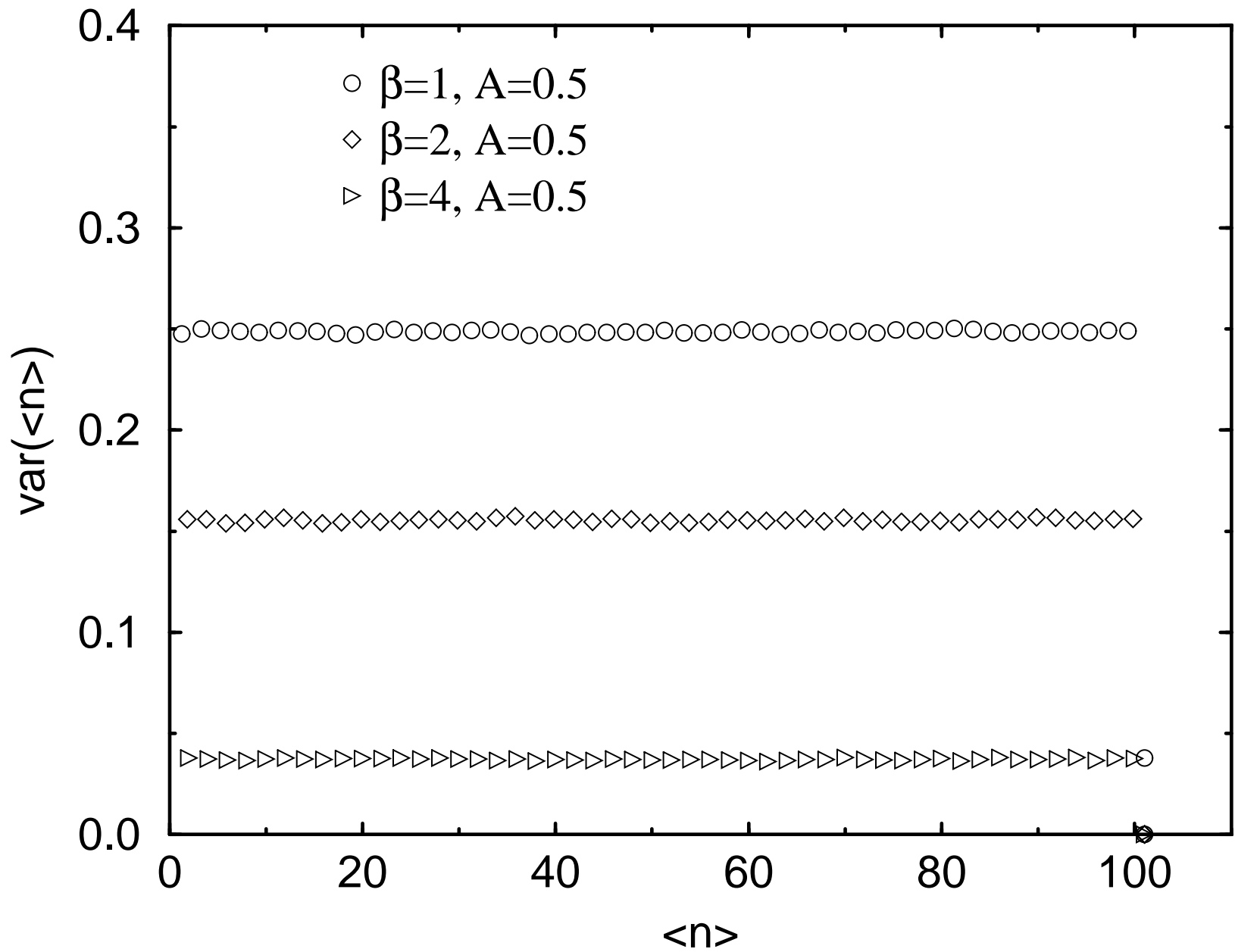
\begin{figure}
\caption{The MC variance vs. $\langle n\rangle$
for the logarithmic potential for
the three symmetries and $A=0.5$. The energy windows are now centered at
the origin. Since the sum rule Eq.~(\protect\ref{sumR})
is satisfied, the linear term in the
variance (see Fig.~\protect\ref{fig8})
vanishes and the system becomes even more rigid than the GE.}
\label{fig9}
\end{figure}

Indeed, the Monte Carlo simulations show
a dramatic difference in the level number variance in these two cases.
In Fig.~\ref{fig8} we show the number variance for an energy window centered
at a point in the bulk, excluding the origin. The variance
grows linearly and the coefficient of the linear term is in good
agreement with Eq.~(\ref{etaD}) and
with the result obtained for the exactly soluble models
by Blecken {\it at al.}.\cite{blecken}
On the other hand, Fig.~\ref{fig9} shows the variance calculated for
symmetric energy windows, containing the origin. The linear term
is absent and the variance is {\it constant} for all integers
$\langle n \rangle >> 1$. Thus, despite the smaller level repulsion, the
overall ``level'' rigidity is even higher
that for the classical RMT.

\section{Discussion and connection with the critical level statistics of the
Anderson model}
\label{sec6}
Let us first summarize the main results of our analysis of the generalized
RMT with soft confinement. We have seen that for the very weak
logarithmic potential,
$V(\epsilon) \sim A\ln^2|\epsilon|$, the local
level fluctuations in the bulk of the spectrum display a cross-over from
the WD to a more Poisson-like behavior, when the parameter $A$ is
decreased. In particular, the
two-level correlation function, in the bulk of the spectrum far from the
origin, decays exponentially at large distances.
The spacing distribution function still vanishes like $s^{\beta}$ at
short separation, but the initial slope is steeper, implying less level
repulsion. The tail of the distribution decays like
$\exp(-a{(A,\beta)}\sigma)$
with $a(A,\beta) >1$, intermediate between the WD surmise and the Poisson
function.
The level number variance, when calculated within
energy windows in the bulk of the spectrum that exclude the origin, is
also Poissonian, increasing linearly with the the average number of levels
$\langle n\rangle$. We saw, however, that the Poisson limit cannot
be reached fully within this ensemble. The nonuniversal behavior
of the RME with logarithmic confining potential has been attributed
to a spontaneous breaking of the $U(N)$ invariance.
For steeper confining potentials,
$V(\epsilon) =|\epsilon|^{\alpha}$, no deviation from the WD
statistics occurs in the bulk of the spectrum, the only small
deviations from universality occurring at the center.\cite{RMpap1}

The question that now we want to address is: does the non classical
(namely non WD) behavior of the RME with double logarithmic confinement
have anything to do with the universal energy-level statistics
of the Anderson model at the critical point? A first point that needs
addressing, before any comparison of the different statistical properties
is attempted,
is the way in which nonuniversality comes about in the invariant RMT
with weak confinement. We have
seen that the essential ingredient to obtain a deviation form the WD
statistics is the strong energy dependence of the averaged level density:
even in the $N\to \infty$ limit, $\langle \rho(\epsilon)\rangle$ is
a rapidly varying function of $\epsilon$ everywhere in the spectrum.
In fact one cannot even define
a constant {\it local } density, since the relative variation of
$\langle \rho(\epsilon)\rangle$ over an energy range equal to the mean
level spacing is of order $1$.
This is at
odds with the well-known result for the density of states in the
Anderson model: in that case $\rho(\epsilon)$ is constant over a large
energy region around the center,
and moreover it is a {\it noncritical} quantity, that is, it does
not exhibit
any drastic change at the critical point. One can reply to this
serious objection by reminding the similarly well-known fact
that quite often complex systems
with different {\it global} statistical properties
(such the density of levels)
have the same {\it local} level fluctuations.\cite{dyson72} and
{\it viceversa}.
The most famous example is the
GE itself, whose semicircle law for the density of states is certainly
not obeyed by any of the spectra of the heavy nuclei or other complex
systems; yet its correlations are very universal and describe accurately
the local statistical properties of these systems. Something of this
sort might happen in our case. In this respect it is
of great importance the existence of another ensemble, namely the broken
symmetric model of Eq.~(\ref{ShE}), where the
level density is constant in the thermodynamic limit,
and nevertheless the local
level correlations are the same of the RME with weak confinement.
Thus there exist at least two RME, having very different
global statistics, whose local statistics belong to the same
new universality class.
We also must emphasize again
that the asymptotic logarithmic behavior for the
confining potential of the invariant RME has been suggested by studies
on transfer matrix models of disordered conductors through the maximum
entropy principle.\cite{Pichardreview,transfermatrix1,transfermatrix2}
In the transfer matrix formalism one can express the conductance $g$ in terms
of the eigenvalues $x_i$ of the matrix
$X = TT^{\dagger} + (TT^{\dagger})^{-1} - 2I$, where $T$ is the transfer
matrix and $I$ the unit matrix
\begin{equation}
g = \sum _{i=1}^N {1\over 1 + x_i}\; .
\end{equation}
Localization appears in the presence of exponentially large eigenvalues $x_i$.
Therefore a simple maximum entropy principle can provide, through the
average density $\rho(x)$, information about the localization of the
wavefunctions, in contrast to what happens to the Hamiltonian matrix.
Here we have assumed that an $U(N)$ invariant RMT for the Hamiltonian matrix
can be constructed from the
confining potential derived from the corresponding transfer matrix.
The hope is that such RMT
will generate the correct local energy-level statistics, despite the
aearge energy density itself will not be well represented.\cite{notemetal}
This procedure is
probably too
naive, but it is clearly the simplest and we will discuss
its implications.

The conjecture of the existence of universal statistical properties
at the metal-insulator transition was put forward by
Shkloskii {\it et al.},\cite{shklovskii}
on the
basis of numerical studies of the spacing distribution function (LSDF) which
turned out to be scale invariant at the critical point.
In a recent work Kravtsov {\it et al.}\cite{KLAA} have carried out
an analytical study of the critical statistics of the Anderson model.
By using the analytical properties of the diffusion propagator and certain
scaling relations valid at the mobility edge they have proved that the
two-level correlation function has the following asymptotic behavior:
\begin{mathletters}
\begin{eqnarray}
\label{r2_crit}
R(s,s') & =& C\beta^{-1}|s-s'|^{-2 +\gamma}\;, \qquad
|s-s'|>>1\; ,\\
\label{gam}
\ &{\phantom =}& \quad \gamma = 1 -(\nu d)^{-1}<1\;,
\end{eqnarray}
\end{mathletters}
where C is a positive constant
while $\gamma$ is an universal critical exponent related
to the critical exponent $\nu$ of the correlation length $\xi$.

However the level number variance at the critical point contains
two terms\cite{ArKL}
\begin{equation}
\label{crit_var}
{\rm var}(\langle n\rangle) =
\eta\langle n\rangle + b\langle n\rangle ^{\gamma}\; ,
\end{equation}
where $\eta<1$ and $b$ are some universal positive constants.
The power-law term originates
directly from the asymptotic power-law tail in the critical two-level
correlator and thus reflects the critical dynamics.
But there is also a {\it linear} term, which had been already
predicted by Altshuler {\it et al.}\cite{altshlinear88}. Formally the origin
of this term is again due to the violation of the sum rule (\ref{etaD}),
by the critical
two-level correlation function.
The physical meaning of this term is not yet
understood. Its existence however implies that the dominant
term in the variance at the critical point is still Poissonian, albei
with coefficient less than 1.
Notice that the knowledge of the two-level correlation function is not
sufficient to develop a complete statistical description of the energy
level at the transition. For example, the LSDF
cannot be found without further statistical assumptions. By mapping
the critical energy levels into a plasma model and
{\it assuming} a the existence of a particular
pair-wise interaction\cite{note3},
one can use the analytical result of Eq.~(\ref{r2_crit})
for the two-level correlation function to derive explicitly the
effective
repulsive interaction among the levels.\cite{KLplasma}.
Once the resulting interaction is
known the asymptotic form of the LSDF can be evaluated,
obtaining the result\cite{KLplasma}
\begin{equation}
\label{P_KLplasma}
P(\sigma) \sim \exp\left(- h_{\gamma}\sigma^{2-\gamma}\right)\; ,
\end{equation}
where $h_{\gamma}$ is a positive constant. Despite some numerical
simulations seem to support this finding (but see below),
this approach has the serious drawback that it gives rise only to the
second term
of Eq.~(\ref{crit_var}) for the level number variance, the linear term
being absent and totally unexplained.

Following the work by Shklovskii {\it et al.}, several other groups
have studied numerically the energy level statistics at the critical point.
In all cases the statistical fluctuation property that is easiest to
study numerically,
namely the LSDF, shows scale invariance and a behavior intermediate
between the WD surmise and the Possion function. There seems to be
agreement also on the linear start of $P(\sigma)$ at small $\sigma$,
with a slope steeper that the WD function for the metallic regime.
However the large $s$ tail behavior is more controversial.
Refs.~[\onlinecite{shklovskii},\onlinecite{zar1}] claim that
$P(\sigma)$ has a Poissonian decay at large $\sigma$
\begin{equation}
P(\sigma) \sim \exp(-a\sigma), \qquad \sigma>>1\;,
\end{equation}
with $a\approx 1.9$, wheras
Refs.~[\onlinecite{evangelo},\onlinecite{hof94}] suggest a behavior
in agreement with the
plasma model result of Eq.~(\ref{P_KLplasma}). We would like
to emphasize that the
numerical results published in Ref.~[\onlinecite{zar1}]
explicitly show good statistics for large values of $\sigma$,
and therefore we believe that they are reliable to extract the
asymptotic behavior
of $P(\sigma)$.

The second important result provided by the numerical simulations
is the existence of a linear term in the variance, as in
Eq.~[\ref{crit_var}].
These
calculations do not exclude the presence of a power-law term of the
kind $\eta\langle n\rangle ^{\gamma}$, which is however difficult to
detect and quantify because of the presence of the dominant
linear term.\cite{sear}.
The coefficient of the linear term is shown to be
$\eta\approx 0.27$ in Ref.~[\onlinecite{zar1}] and $0.30$ in
Ref.~[\onlinecite{sear}].

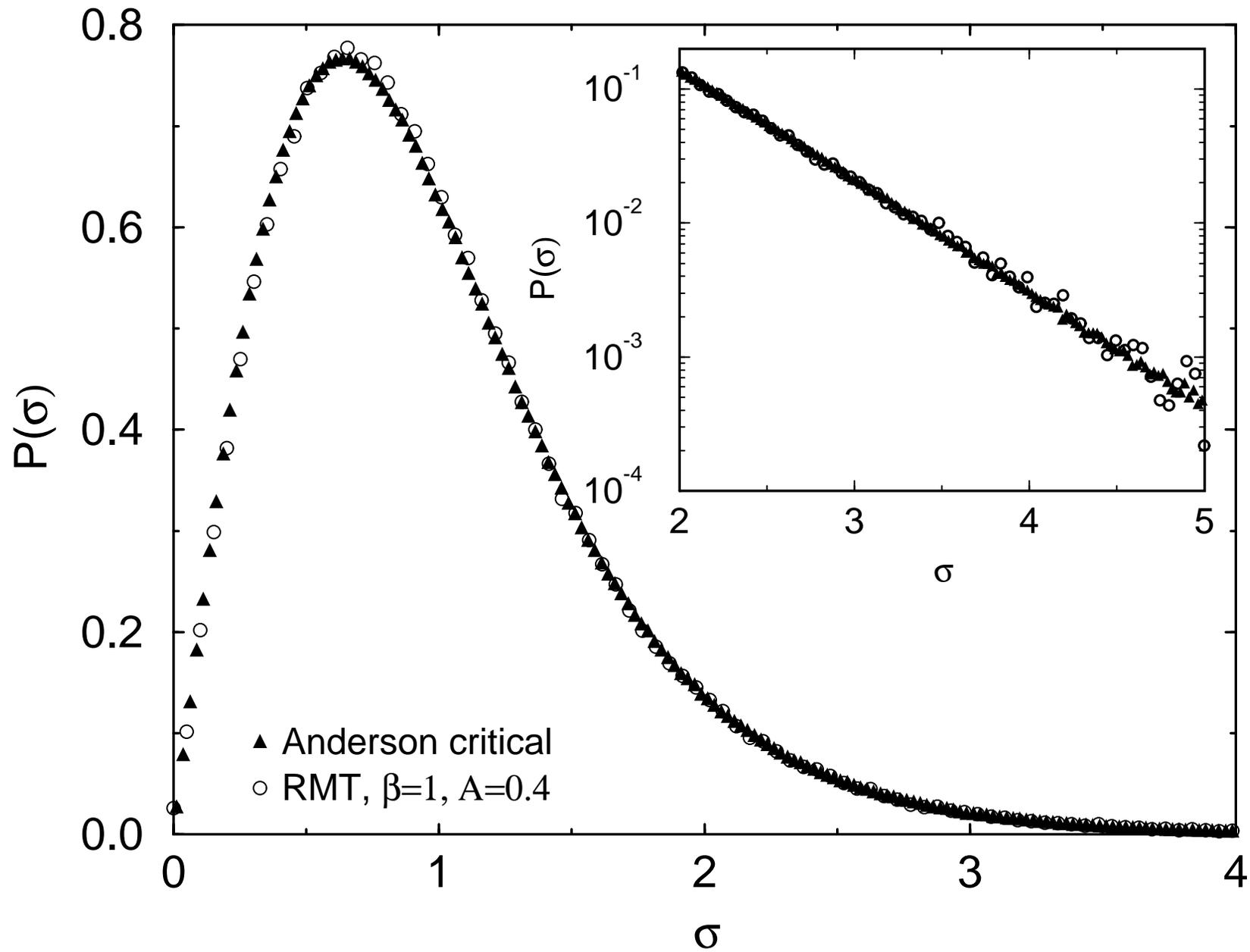
\begin{figure}
\caption{LSDF of the RME with logarithmic confinement for
$\beta =1$, $A=0.4$,
plotted together with the energy-LSDF of the three dimensional Anderson model
at the metal-insulator transition, taken from
Ref.~[\protect\onlinecite{zar1}].
The inset shows the large $\sigma$ behavior in logarithmic scale for the
$y$-axis. A straight line is the best fit for both curves.}
\label{fig10}
\end{figure}

The RME with weak confinement that we have considered in this paper is able
to reproduce
two of the main features
seen in the numerical simulations of the critical
statistics: the overall behavior of $P(\sigma)$--
with the linear rise at $\sigma<<1$ and the exponential decay
at $\sigma>>1$)--
and the linear dependence of the number variance.
In Fig.~\ref{fig10} we plot the LSDF of the RME with logarithmic
potential for
$A=0.4, \beta=1$, together with the critical LSDF of the Anderson model
from Ref.~[\onlinecite{zar1}]. The agreement between the two
curves is spectacular in a very large energy range where $P(\sigma)$
varies by five orders of magnitude.
Notice in the inset of the figure the behavior
for large $\sigma$ of the tail of $\ln[P(\sigma)]$,
which has apparently a linear slope.
The parameter $A_c\approx 0.4$
identifies, among all the possible members
of the family of RME with logarithmic confinement,
the ensemble which has the closest LSDF to the critical statistics.

If we now compute the coefficient $\eta$ of the linear term in the variance
for the RME with $A_c\approx 0.4$ [see Fig.~{fig8}],
we obtain $\eta\approx 0.32$,
which is
consistent with the numerical results from the exact
diagonalizations.\cite{sear,zar1}
Thus the RME with logarithmic
confinement is able to reproduce quantitatively
the shape of the critical LSDF when $A=A_c$ and at the same time provides
an accurate estimate of the
the leading order term of the number variance. In some sense, the
``residual Possonian'' properties of the critical statistics are
well reproduced by this generalized RMT.
The RMT does not provide the asymptotic power-law behavior
(\ref{r2_crit}) of the
two-level correlation function found analytically, which,
on the other hand, is also difficult to extract by direct numerical
diagonalizations with reliable accuracy.\cite{montanbaux}
Clearly more work is necessary to determine if the good
agreement shown here between the RME with weak confinement and
the critical statistics is more than a furtuitous coincidence.
It is however interesting and important that some of the
properties of the correlations
at the mobility edge can be correctly reproduced by such a simple
ensemble.
\section{Conclusions}
\label{sec7}
In this paper we have studied in detail the properties of families of
random matrix ensembles (RME) that are
invariant under similarity transformations but are characterized by
a generalized level confinement.
We have shown that the level statistics
are affected by the confining potential when this is very soft.
In particular, for a squared logarithmic potential, the statistical bulk
properties are nonuniversal and deviate significantly
from the Wigner-Dyson statistics of the Gaussian Ensembles,
exhibiting a cross-over
toward a more Poissonian behavior when an internal parameter is
decreased. The $U(N)$ invariant RME with logarithmic confining
potential belongs,
together the RME with a symmetry breaking term [ {\it see} Eq.~(\ref{ShE})],
to a new universality class, distinct from
Wigner-Dyson universality of classical RMT.

We have shown that the nonuniversal behavior of
the two-level correlation function for these RME can still be
obtained within
Dyson's mean field theory, generalized to the case weak confinement.

We have performed Monte Carlo simulations to calculate several important
statistical properties of the generalized RME that probe both short
and long range correlations.

The statistical properties of the RME with logarithmic confinement
have strong similarities with the universal energy level statistics
of disordered conductors at the metal-insulator transition.
In particular, the probability distribution of the level spacings
for a three dimensional Anderson model at the critical point can
be very well fitted, throughout a wide energy range,
by the corresponding RME function for one
particular choice of the internal parameter. Then for the {\it same} value
of the
parameter, this RME predicts a linear behavior for the level number variance,
with a coefficient of proportionality close to the value obtained from
numerical diagonalizations of the Anderson model at the critical point.

\acknowledgments

I am very grateful to V.~E.~Kravtsov for introducing me to this field
and for providing important suggestions and advise throughout the stage of
this work.
I also want to thank Mats~Wallin and Yu~Lu of useful discussions,
and I.~Kh.~Zharekeshev for sending me the data file with the results published
in Ref.~[\onlinecite{zar1}].


\begin{references}
\bibitem{note0} For an insulator in the $L\to\infty$ limit, the
        energy level statistics are universal and exactly equal
        to the Poisson distribution, since localized electron states
        corresponding to different energies do not overlap in space
        and therefore the energy levels are uncorrelated.
\bibitem{note00} The singularity at $\epsilon=0$ is irrelevant since it
        does not affect the bulk properties. Other potentials, differently
        regularized at the origin but with the same asymptotic
        behavior will have the same statistical properties in the bulk.
\bibitem{noteonmu} We have again dropped the term with the
        functional derivative of $\mu$. Such an approximation is admittedly
        less justified in this case, since, even if the density is now
        constant in the bulk, the correlations are not. However
        what follows from neglecting this term shows that the
        nonconstant density plays the major role here and is
        responsible for setting up the nonuniversal behavior.
\bibitem{noteSUMR} Notice that the GE satisfies the sum rule, while
         in the case of Poisson statistics the sum rule is
         maximally violated : $\eta =1$.
\bibitem{note1} We will use the language taken from the plasma
        model and call ``particles'' the energy levels.
\bibitem{noteR2} To compute $R_2(0,s')$, the conditional density
        must then be normalized by the normal density $\rho(\epsilon')$
        computed numerically:
        $R_2(0,s') = R_2(0,\epsilon_{s'})/\rho(\epsilon_{s'})$.
\bibitem{note2} This form is for the Orthogonal ensemble ($\beta=1$).
        For the other ensembles one can similarly define
        ``Wigner surmises'' of
        the form $P(\sigma) = c_1 \sigma^{\beta}\exp(- c_2 \sigma^2)$ where
        the constants $c_1$ and $c_2$
        depend on $\beta$ and are found by imposing the conditions
        that $P(\sigma)$ and its first moment are equal to 1.%
\bibitem{notemetal} Notice that in the {\it metallic regime} the
        local statistics
        of the Hamiltonian eigenvalues and the $X$-matrix eigenvalues
        are equal, being in both cases equal to the universal WD statistics.
        It is not obvious that the same is true at the metal-insulator
        transition.
\bibitem{note3} The main assumption is to take an effective pair-wise
        interaction of the form
        $f(\epsilon-\epsilon') \sim |\epsilon-\epsilon'|^{\delta}$, with
        ${\delta}<1$. This allows to use a MFT treatment to find
        ${\delta}$ from the asymptotic form of $R_2(s)$, given in
        Eq.~(\ref{r2_crit}), yielding
        $\delta = \gamma= 1 -1/(\nu d)$.
\end{references}
\end{document}